\documentclass[showpacs,superscriptaddress]{revtex4}%%%%%
\usepackage{booktabs}
\usepackage[center,small,it]{subfigure} %%%%%%%
\usepackage{doublespace,amsmath,amsfonts,graphicx} %%%%
\usepackage{booktabs} %%%%%%%%%%%%%%%%
\begin{document}%%%%%%%%%%%%%%%%%%%
\begin{spacing}{1.1}%%%%%%%%%%%%%%%%%
%%%%%%%%%%%%%%%%%%%%%%%%%%%%%%%%%%%%%%%%%%%%
%%
% Última corrección, 22 de marzo de 2002
%%
%%%%%%%%%%%%%%%%%%%%%%%%%%%%%%%%%%%%%%%%%%%%
\title{Hierarchical radiative fermion masses and flavor changing processes
in an $U(1)_X$ horizontal symmetry model} 
\author{E. Garc\'{\i}a-Reyes}
\affiliation{Departamento de F\'{\i}sica, Cinvestav \\ 
 Apartado Postal 14-740, 07000 \\ 
 M\'exico, D.F., M\'exico.}

\author{A. Hern\'andez-Galeana}
\affiliation{Departamento de F\'{\i}sica, Escuela Superior de F\'{\i}sica y 
Matem\'aticas \\ 
Instituto Polit\'ecnico Nacional, U. P. Adolfo L\'opez Mateos \\ 
M\'exico D.F., 07738, M\'exico } %

\author{A. Vargas}
\author{A. Zepeda}
\affiliation{Departamento de F\'{\i}sica, Cinvestav \\ 
 Apartado Postal 14-740, 07000 \\ 
 M\'exico, D.F., M\'exico.}

%\date{} 

\pacs{12.15.Ff, 12.10.-g}

% Última revisión, 22 de Marzo de 2002
%

\begin{abstract}
In a model with a gauge group $G_{SM}\otimes U(1)_X$, where $G_{SM} \equiv 
SU(3)_C \otimes SU(2)_L \otimes U(1)_Y$ is the standard model gauge group 
and $U(1)_X$ is a horizontal local gauge symmetry,  the spectrum of charged 
fermion masses and mixing angles are obtained from radiative contributions 
with the aim to reproduce the fermion masses and the Cabibbo-Kobayashi-Maskawa
 matrix with no {\it a priori} texture assumption. The assignment of 
horizontal charges is made such that at tree level only the third family 
becomes massive when electroweak symmetry is broken. Using these tree level 
masses and introducing exotic scalars, the light families of charged fermions 
acquire hierarchical masses through radiative corrections. The  rank three 
mass matrices obtained are written in terms of a minimal set of free 
parameters in order to estimate their values by  a numerical fit performance.
 The resulting  masses and CKM mixing angles turn out to be in good agreement
 with the known values.

The new exotic scalar fields mediate flavor changing processes and contribute
 to the anomalous magnetic moments of fermions, additionaly branching ratios and anomalous
 magnetic moments contributions were  evaluated for leptons.

\end{abstract}

\maketitle

%$$$$$$$$$$$$$$$$$$$$$$$$$$$$$$$$$$$$$$
%Última corrección, 22 de marzo de 2002
%
\section{INTRODUCTION} 
In this work a continuous horizontal local symmetry $U(1)_X$ is considered.
%, which is spontaneously broken, and 
Assuming that only the third charged fermion family becomes massive at tree level, the light fermion masses are produced radiatively, generated at one and two loops. Instead of assuming a specific texture %from the beginning 
for the mass 
matrices, explicit one and two loop contributions are calculated and the 
mass matrix texture are obtained from these calculations. The textures arise because of the scalar field content of the model, which in turn are a 
consequence of the required gauge symmetry. 

The new scalar fields needed to produce the light fermion masses mediate
flavor changing processes and contribute to the anomalous magnetic moment
of the fermions. The contributions to the anomalous magnetic moment and flavor
changing processes are drawn for the leptons to show that no experimental
bound is violated.

So, the radiative generation of hierarchical masses is possible within a
model which do not need new exotic fermions and, at the same time is
consistent with the experimental bounds on rare processes. Even though the
$U(1)_X$ breaking scale is not known the scalar masses which arise from the numerical
fit are not in contradiction with well known estimations on the scale of
new physics, and the scalar contributions to rare decays and anomalous
magnetic moments still allow for the search of new gauge boson effects.

The analysis of the model in question to generate the masses and mixing angles of the quark sector is already reorted in \cite{Masas-RMF}

This paper is organized in the following way: In section II an
explicit description of the model is given, section III presents the 
analytical calculations for fermion masses and later the scalar contributions to lepton anomalous magnetic moment and rare decay branching ratios, while in section IV the numerical fit results
are shown. Comments and conclusions are presented in section V.
% Última corrección, 22 de marzo de 2002
\section{The Model}

Only three fermion families are assumed, those of the Standard Model
(SM), and then no exotic fermions need to be introduced to cancel
anomalies. The fermions are classified, as in the SM, in five sectors
\textit{f} = \textit{q,u,d,l} and \textit{e}, where \textit{q} and
\textit{l} are the $SU(2)_L$ quark and lepton doublets respectively and
\textit{u,d} and \textit{e} are the singlets, in an obvious notation.  In
order to reduce the number of parameters and to make the model free of
anomalies, the values X of the horizontal charge are demanded to satisfy
the traceless condition\cite{ZEP1}
\begin{equation}\label{Eq. 1}
X(f_i)=0,\pm \delta_f,
\end{equation}
where $i=1,2,3$ is a family index, with the constraint 
\begin{equation}\label{Eq. 2}
\delta_q^2 - 2 \delta_u^2 + \delta_d^2 = \delta_l^2 - \delta_e^2.
\end{equation}
Eq. (\ref{Eq. 1}) guarantees the cancellation of the [U(1)$_X$]$^3$
anomaly as well as those which are linear in the U(1)$_X$ hypercharge
($[SU(3)_C]^2U(1)_X,\; [SU(2)_L]^2U(1)_X,\; [Grav]^2U(1)_X$ and
$[U(1)_Y]^2U(1)_X$). Eq. (\ref{Eq. 2})
is the condition for the cancellation of the U(1)$_Y$[U(1)$_X$]$^2$
anomaly. A solution of Eq. (\ref{Eq. 2}) which guarantees
that only the top and bottom quarks acquire masses at tree level is given 
by
(``doublets independent of singlets", see Ref. \cite{ZEP1})  
\begin{equation}\label{Eq. 3}
 \delta_l = \delta_q = \Delta \neq \delta_u = \delta_d = \delta_e
=\delta.
\end{equation}
%\vskip.3cm
To  avoid tree level flavor changing neutral currents mediated by the standard Z boson, the
mixing between the standard model Z boson and its horizontal counterpart 
is not allowed.
Consequently the SM Higgs scalar should have zero
horizontal charge. As a consequence, and since a
non-zero tree-level mass
for the top and bottom quarks is required, the horizontal charges of these 
quarks 
should satisfy
\begin{eqnarray}
-X(q_3) + X(u_3) = 0 \mbox{\hspace{1cm}and\hspace{1cm}} & -X(q_3) + X(d_3) = 0
\end{eqnarray}
in order for the Yukawa couplings in Eq. (\ref{pyc}) to be invariant,
but then Eqs. (\ref{Eq. 1}) and (\ref{Eq. 3}) demand that they vanish,
\begin{equation}\label{Eq. 4}
 X(u_{3}) = X(q_{3}) = X(d_{3}) = 0,
\end{equation}
which in turn implies $X(l_{3})=X(e_{3})=0$ (this defines the third family). 
The assignment of horizontal charges to the fermions is then as given in
Table 1. The  $SU(3)_C \otimes SU(2)_L \otimes U(1)_Y$ quantum numbers of
the fermions are the same as in the Standard Model.

\begin{table} \centering
\subtable[Scalars\label{Esca:Def}]{%
\begin{tabular}{ccc} \\\toprule
  &\em Class I & \em \hspace{1.5cm}Class II  \hspace{1.2cm} Class I \hspace{2mm}  Class II\\ \cmidrule{2-3}
\hline
\begin{tabular}{c} \\\cmidrule{1-1} X \\  Y \\ 
T \\C \end{tabular} &
\begin{tabular}{cc} $\phi_1$& $\phi_2$ \\\cmidrule{1-2}
                      0 & $-\delta$ \\ 1 & 0 \\ $\frac{1}{2}$ & 0  \\ 1 & 1 \\\end{tabular} &
\begin{tabular}{cccccc|cc|cc} 
$\phi_3$&$\phi_4$&$\phi_5$&$\phi_6$&$\phi_7$&$\phi_8$&$\phi_9$&$\phi_{10}$&$\phi_{11}$&$\phi_{12}$ \\\cmidrule{1-10}
 0 & $\Delta$ & 0 & $\delta$ & 0 & $\delta$ &$\Delta$& 0& $\delta$& 0 \\ 
 $-\frac{2}{3}$& $-\frac{2}{3}$ &$\frac{4}{3}$& $\frac{4}{3}$&$-\frac{8}{3}$& $-\frac{8}{3}$ & 2& 2& 4& 4 \\ 
 1 &1 & 0 & 0 & 0 & 0 &1& 1& 0& 0\\ 
 $\bar{6}$ & $\bar{6}$ & $\bar{6}$ & $\bar{6}$ & $\bar{6}$ & $\bar{6}$ &  1 &1& 1 &1 \end{tabular}\\\bottomrule
\end{tabular}}
\hspace{1cm}
%\\
%%%%%%%%%%
%\begin{floatingtable}[l]{
\subtable[Fermions\label{Fermiones:NC}]{%
\begin{tabular}{c|ccc} \toprule
        &\multicolumn{3}{c}{Family}\\\hline%\cmidrule{2-4}
 Sector &    1    &     2    & 3 \\ \midrule
 $q$             & $\Delta$ & $-\Delta$ & 0 \\
 $u$             & $\delta$ & $-\delta$ & 0 \\
 $d$             & $\delta$ & $-\delta$ & 0 \\
 $l$             & $\Delta$ & $-\Delta$ & 0 \\
 $e$             & $\delta$ & $-\delta$ & 0 \\ \bottomrule
\end{tabular}}
\label{tablas}
\caption{Quantum numbers for the scalar fields and fermions horizontal 
charges.}
\end{table}

To generate the first and second family charged fermion masses radiatively 
 new irreducible representations (irreps) of scalar fields should be 
introduced, 
since the gauge bosons of $G=G_{SM}\otimes U(1)_X$ do not perform transitions 
between different families. Families are of course distinguishable (non 
degenerated), only below the scale of
the electroweak symmetry breaking, when they become massive.

Looking for scalars which make possible the generation of fermion masses 
in a hierarchical manner, the irreps  of scalar fields are divided into 
two classes. Class I (II) contains scalar fields which acquire (do not acquire) vacuum expectation value (VEV). 

A suitable choice of scalars should be made in order to avoid induced VEVs, through couplings in the potential, for class II scalars. In the model considered below class II scalars have no electrically neutral components, so they never get out of their class. In this model four irreps of scalars of class I and eight irreps of scalars of class II are introduced, with the quantum numbers  specified in Table \ref{Esca:Def}. Notice that just the minimum number of class I scalars are introduced; i.e., only one Higgs weak isospin doublet to achieve Spontaneous Symmetry Breaking (SSB) of the electroweak group down to the electromagnetic $U(1)_Q$, one $SU(2)_L$ singlet $\phi_2$ used to break  $U(1)_X$, and two weak isospin triplets ($\phi_9$ and $\phi_10$) which produce very small contributions to the electroweak SSB. In this way the horizontal interactions affect the $\rho$ parameter only at higher orders.

With the above quantum numbers the Yukawa couplings that can be written 
may be divided into two classes, those of the D type which are defined by 
Fig. \ref{Gen:a}, and  those of the M  type which are defined in 
Fig. \ref{Gen:b}. The Yukawa couplings can thus be written as $L_{Y}
= L_{Y_D} + L_{Y_M}$, where the D Yukawa couplings are
\begin{equation}\label{pyc}
L_{YD}  =  Y^{u} \bar{q}_{L3} \tilde{\phi}_{1} u_{R3} + Y^{d} \bar{q}_{L3} \phi_{1} d_{R3}  + Y^{\tau} \bar{l}_{3L} \phi_1 \tau_R  + h.c.,
\end{equation}
with $\tilde{\phi}\equiv i\sigma_2\phi^*$, while the M couplings
compatible with the symmetries of the model are
\begin{eqnarray}\label{iyc}
L_{YM}=Y_{Q}[q_{1L}^{\alpha T} C \phi_{3 \{ \alpha \beta \} }
q_{2L}^{\beta} + q_{3L}^{\alpha T} C \phi_{3 \{ \alpha \beta \} }
q_{3L}^{\beta} + q_{2L}^{\alpha T}
C \phi_{4 \{ \alpha \beta \} } q_{3L}^{\beta} \nonumber\\
+ d_{2R}^{T} C
\phi_5 d_{1R} + d_{3R}^{T} C \phi_5
d_{3R} + d_{3R}^{T} C \phi_6 d_{2R} %\nonumber\\
+ u_{2R}^{T} C \phi_7 u_{1R} +
u_{3R}^{T} C
\phi_7 u_{3R} + u_{3R}^{T} C \phi_8 u_{2R}] \nonumber \\+
 Y_l [ l_{2L}^{\alpha T} C \phi_{9 \{ \alpha \beta \} } l_{3L}^{\beta} + l_{1L}^{\alpha T} C \phi_{10 \{ \alpha \beta \} } l_{2L}^{\beta} + l_{3L}^{\alpha T} C \phi_{10 \{ \alpha \beta \} } l_{3L}^{\beta} + \mu_R^{T} C\phi_{11}\tau_R + 
\nonumber \\ e_R^{T} C \phi_{12} \mu_R + \tau_R^{T} C \phi_{12} \tau_R ] + h.c.
\end{eqnarray}
In these couplings C represents the charge conjugation matrix and
$\alpha$ and $\beta$ are weak isospin indices. Color indices
have not been written explicitly. By simplicity and economy 
 only one Yukawa
constant $Y_Q$ is assumed for all the quark M couplings and, another one, $Y_l$ for the lepton sector. Notice that
$\phi_{3\{\alpha\beta\}}$ and $\phi_{9\{\alpha\beta\}}$ are represented as

\begin{equation}\label{triplet}
\phi_3 = \left( \begin{array}{cc}
\phi^{-4/3} &  \phi^{-1/3} \\
\phi^{-1/3} & \phi^{2/3} \\
\end{array} \right) 
\text{ and }
\phi_9 = \left( \begin{array}{cc}
\phi^{0} &  \phi^{+} \\
\phi^{+} & \phi^{++} \\
\end{array} \right),
\end{equation}
where the superscript denotes the electric charge of the field (and corresponding expressions for $\phi_4$ and $\phi_{10}$).
%The same applies for the weak isospin triplet $\phi_4$, 
%While $\phi_9$ (and $\phi_{10}$) is represented as
%\begin{equation}\label{triplet-l}
%\phi_9 = \left( \begin{array}{cc}
%\phi^{0} &  \phi^{+} \\
%\phi^{+} & \phi^{++} \\
%\end{array} \right). 
%\end{equation} 
%and similarly for $\phi_{10}$.

Scalar fields which are not SU(2)$_L$ doublets do not participate in D type
Yukawa terms, they however contribute to the mass matrix of the scalar
sector and in turn determine the magnitude of the radiatively generated
masses of fermions, as is shown below.

The most general scalar potential of dimension $\leq 4$ that can be written is 
\begin{eqnarray}\label{Pot4d}
-V(\phi_{i})=%&
 \sum_{i} \mu_{i}^{2} \|\phi_{i}\|^{2} + \sum_{i,j}\lambda_{ij}
\|\phi_{i}\|^{2}\|\phi_{j}\|^{2} + \eta_{31} \phi_{1}^{\dag}
\phi^{\dag}_{3} \phi_{3} \phi_{1} + %\nonumber \\ &
\tilde{\eta_{31}} \tilde{\phi_{1}}^{\dag} 
\phi^{\dag}_{3} \phi_{3} \tilde{\phi_{1}}  \nonumber \\ + 
\eta_{41} \phi_{1}^{\dag} \phi^{\dag}_{4} \phi_{4} \phi_{1} +
\tilde{\eta_{41}} 
\tilde{\phi_{1}}^{\dag} \phi^{\dag}_{4} \phi_{4} \tilde{\phi_{1}}
+ 
\kappa_{9 1} \phi_1^{\dag} 
\phi_9^{\dag} \phi_9 \phi_1 + %\nonumber \\ &
\tilde{\kappa}_{9 1} \tilde{\phi_1}^{\dag} 
\phi_9^{\dag} \phi_9 \tilde{\phi_1} \nonumber \\+ \kappa_{10, 1} \phi_1^{\dag} 
\phi_{10}^{\dag} \phi_{10} \phi_1  + \tilde{\kappa_{10, 1}} 
\tilde{\phi_1}^{\dag} \phi_{10}^{\dag} \phi_{10} \tilde{\phi_1} +
\sum_{ \substack{ i\neq j \\i,j\neq 1,2 }} \eta_{ij} \| \phi_{i}^{\dag} 
\phi_{j} \|^{2}  + %\nonumber \\&
 (\rho_{1} \phi_{5}^{\dag}\phi_{6} \phi_{2} + \nonumber \\ 
\rho_{2} \phi_{7}^{\dag} \phi_{8} \phi_{2}  + \lambda_{1} \phi_{5}^{\dag} \phi_{1}^{\alpha} 
\phi_{3 \{ \alpha \beta \} }\phi_{1}^{ \beta } +  \lambda_{2} \phi_{7}^{\dag} 
\tilde{\phi_{1}}^{\alpha} \phi_{3 \{ \alpha \beta \} }\tilde{\phi_1}^{\beta}  + %\nonumber \\&
 \lambda_3 Tr(\phi_3^{\dag} \phi_4) \phi_2^{2} + \nonumber \\ \lambda_4 \phi_5 \phi_6 
\phi_7 \phi_2 + \lambda_5 \phi_5 \phi_6^{\dag} \phi_7^{\dag} \phi_8 + 
%\lambda_6 \phi_2 \phi_8 \phi_5^{2}   + %\nonumber \\ &
 y_l \phi_{12}^{\dag} \phi_{11}\phi_2 + \nonumber \\ \zeta_1 \phi_{12}^{\dag}
\phi_1^{\alpha} \phi_{10  \{ \alpha , \beta \} } \phi_1^{\beta}  + 
Y_r Tr\left(\phi_{10}^{\dag} \phi_9 \right) \phi_2^{2} + \varepsilon_1 
\phi_5 \phi_6^{\dag} \phi_{12}^{\dag} \phi_{11} + %\nonumber \\&
 \varepsilon_2 
\phi_7^{\dag} \phi_8 \phi_{12} \phi_{11}^{\dag} + h.c.),
\end{eqnarray}
where $Tr$ means trace and in $ \mid
\phi_i \mid ^2  \equiv \phi_i^{\dag} \phi_i $ an appropriate contraction of
the $SU(2)_L$ and $SU(3)_C$ indices is understood. The gauge invariance of
this potential requires the relation $\Delta = 2\delta$ to be hold.

The VEVs of the class I scalar fields are
\begin{eqnarray}\label{vevs}
\langle\phi_1\rangle=\frac{1}{\sqrt{2}}\left(\begin{array}{c} 0 \\ v_1 \end{array} \right),&&\langle\phi_2\rangle=v_2,\nonumber\\
\langle\phi_9\rangle=\left(\begin{array}{cc} v_9 & 0 \\ 0& 0\end{array}\right)&\mbox{and} & \langle\phi_{10}\rangle=\left(\begin{array}{cc} v_{10} & 0 \\ 0& 0\end{array}\right),
\end{eqnarray}
$\langle \phi_1 \rangle$ and $\langle \phi_2 \rangle$ achieve  the symmetry breaking sequence
\begin{equation}\label{cadena}
G_{SM}\otimes U(1)_X \stackrel{<\phi_2>}{\longrightarrow} G_{SM}
\stackrel{<\phi_1>}{\longrightarrow} SU(3)_C \otimes
U(1)_Q,
\end{equation}
while
the VEVs $v_9$ and $v_{10}$ are extremely small and allow 
production of Majorana type mass contributions to the neutrino mass matrix. With these contributions the 3 left handed neutrinos acquire masses, whose differences are consistent with recent analysis \cite{NusBib} and there is no need to introduce right handed neutrinos.
In this case the mass matrix has the following form
\begin{equation}\label{neutrino-mass}
\phi_3 = \left( \begin{array}{ccc}
0 &  v_{10}/2 & 0\\
v_{10}/2 & 0 & v_9/2 \\
0 & v_9/2 & v_10 \\
\end{array} \right) 
\end{equation}

The scalar field mixing arises after SSB  from the terms in the potential that couple two different class II fields to one class I field. After SSB the mass matrix for the scalar fields of charge $2/3$ ($\phi_{4}$,$\phi_{3}$, $\phi_5$,$\phi_6$) %and $4/3$ ($\phi_{4}$,$\phi_{3}$,$\phi_7$,$\phi_8$) are 
is written as%, respectively, as
\begin{eqnarray}\label{scalarmass}
M_{2/3}^2=\left( \begin{array}{cccc}
s_{4}^2   &  \lambda_3^* v_2^2  &      0         &   0 \\
\lambda_3 v_2^2 &  s_{3}^2      & \frac{ \lambda_1^*
v_1^2}{2} & 0 \\
0  & \frac{ \lambda_1 v_1^2}{2} & u_5^2 & \rho_1
v_2 \\
0 & 0 & \rho_1^* v_2 & u_6^2 
\end{array} \right) %& \mbox{and} & M_{4/3}^2=\left(
%\begin{array}{cccc}
%t_{4 }^2   &  \lambda_3^* v_2^2  &      0         &   0 \\ 
%\lambda_3 v_2^2 &  t_{3 }^2  & \frac{ \lambda_2
%v_1^2}{2} & 0 \\ 
%0  & \frac{ \lambda_2^* v_1^2}{2} & t_7^2 & \rho_2
%v_2 \\ 
%0 & 0 & \rho_2^* v_2 & t_8^2 
%\end{array} \right),
\end{eqnarray}
and analogous ones for the $4/3$ and $-1$ charge sectors. 
%From Eq. (\ref{Pot4d}) $t_{i}^2 = u_{i}^2 =
%\mu_i^2+\lambda_{i1}v_1^2+\lambda_{i2}v_2^2 $  
%and $s_{i}^2=t_{i}^2+\eta_{i1}v_1^2$.
%%%%%%%%%%%%%%%%%%%%%%%%%%%%%%%%%%%%%%
%%%%%%%%%%%%%%%%%%%%%%%%%%%%%%%%%%%%%
\section{Analytical Results}
\subsection{The Masses}
A description of the mechanism to produce the charged fermion masses 
follows.

\begin{figure}
\begin{center}
\subfigure[]{%\centering
\includegraphics[height=2cm]{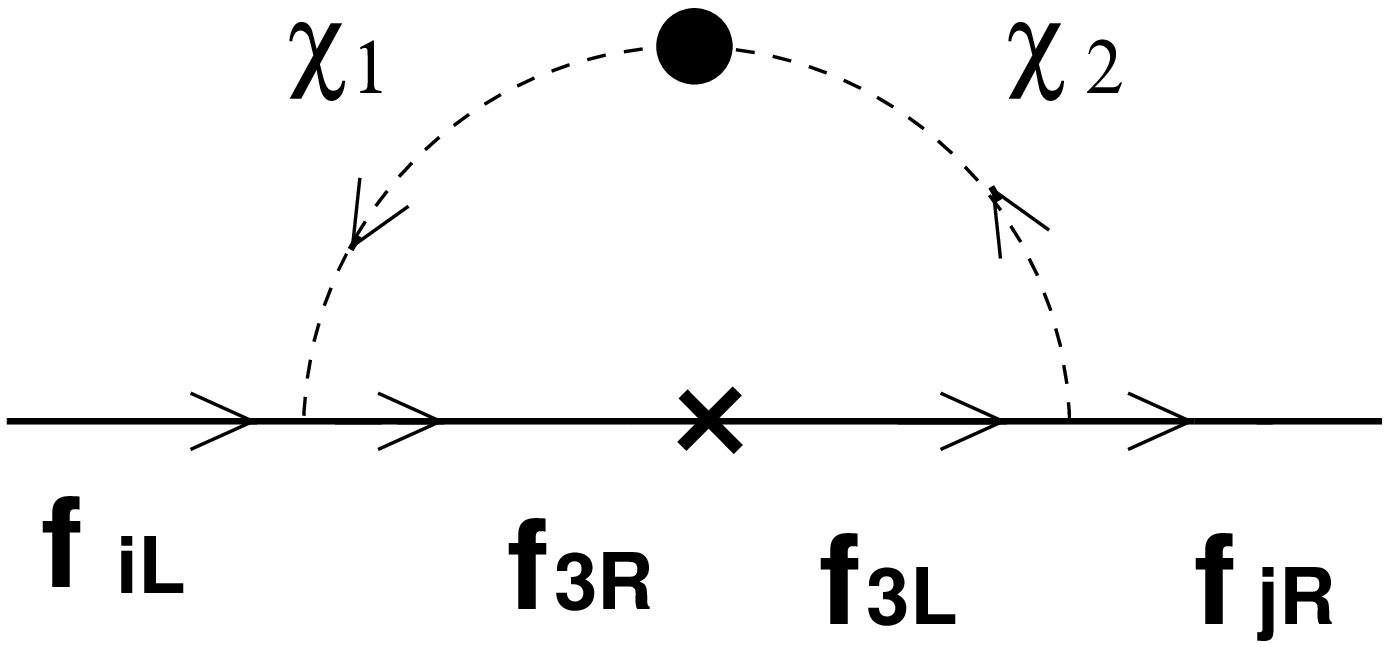}\label{Gen:a}}
\hspace{1.5cm}
\subfigure[]{%\centering
\includegraphics[height=2cm]{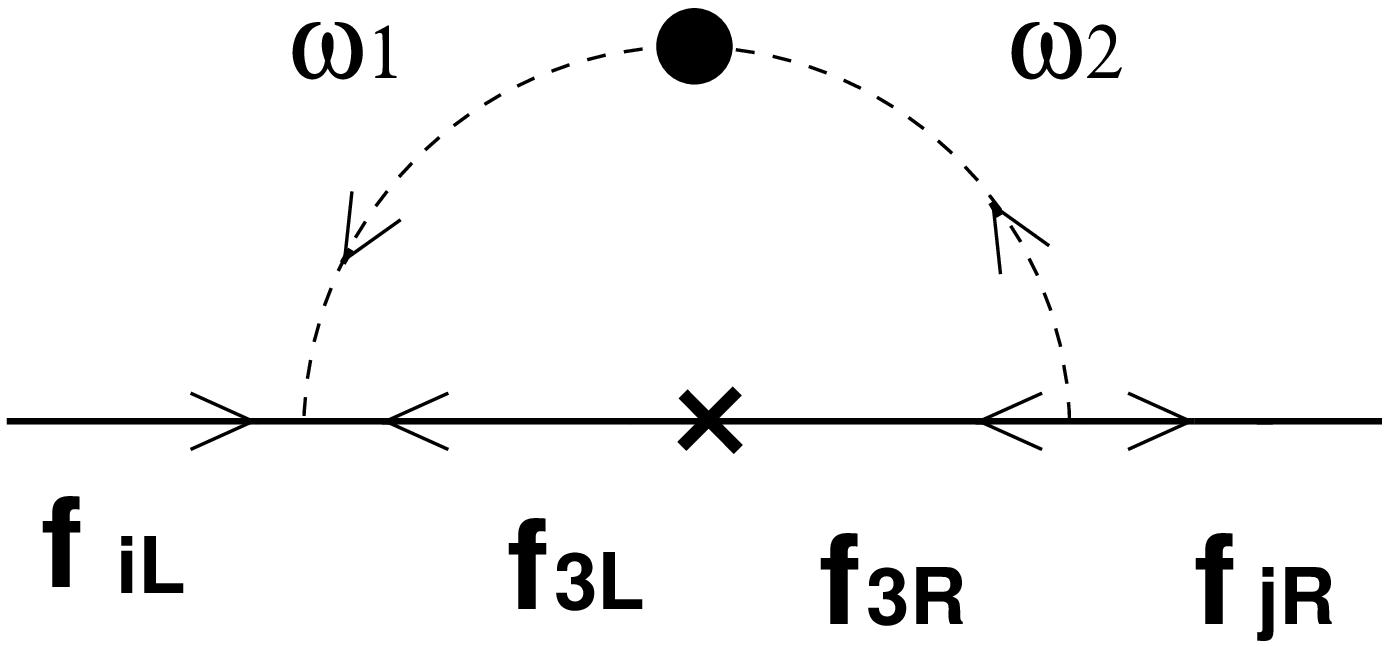}\label{Gen:b}}
\caption{Generic diagrams that could contribute to the mass of the light
families, (a) D type couplings are represented with vertices where
one fermion is incoming and the other one is outgoing, (b) the M    
type couplings are represented with vertices where both fermions are
incoming or outgoing.} \label{M2fig} 
\end{center}
\end{figure}

\begin{figure}[ht]
\centering
\begin{tabular}{ccc}
\includegraphics[height=1.6cm]{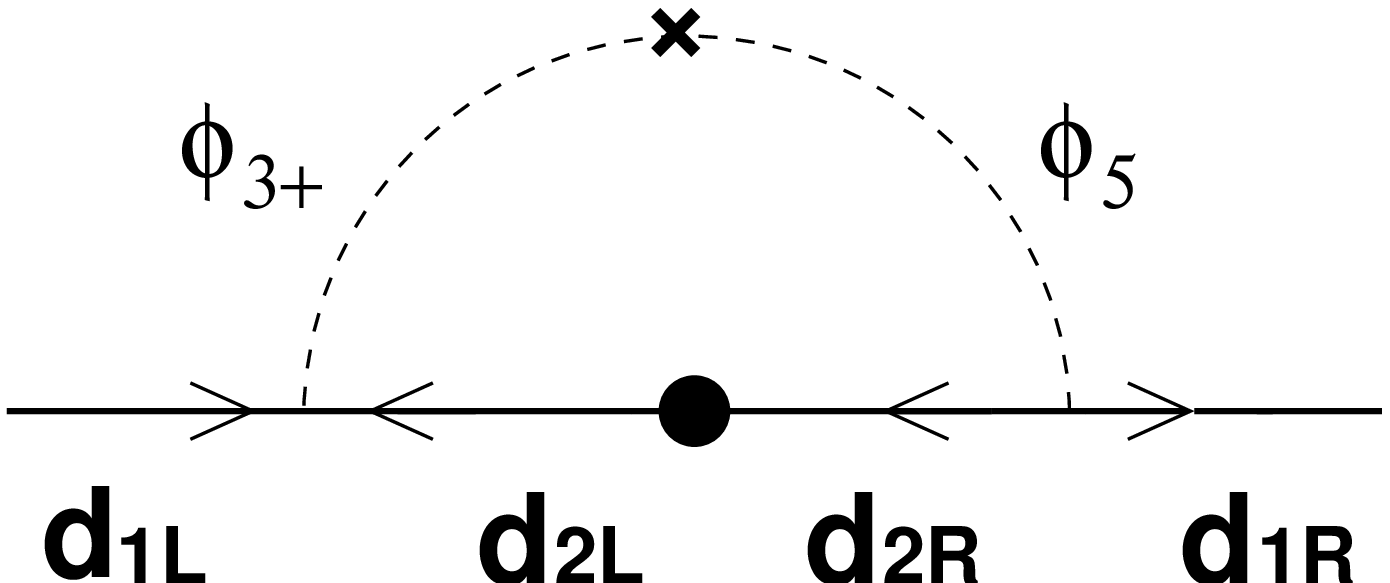} &
\includegraphics[height=1.8cm]{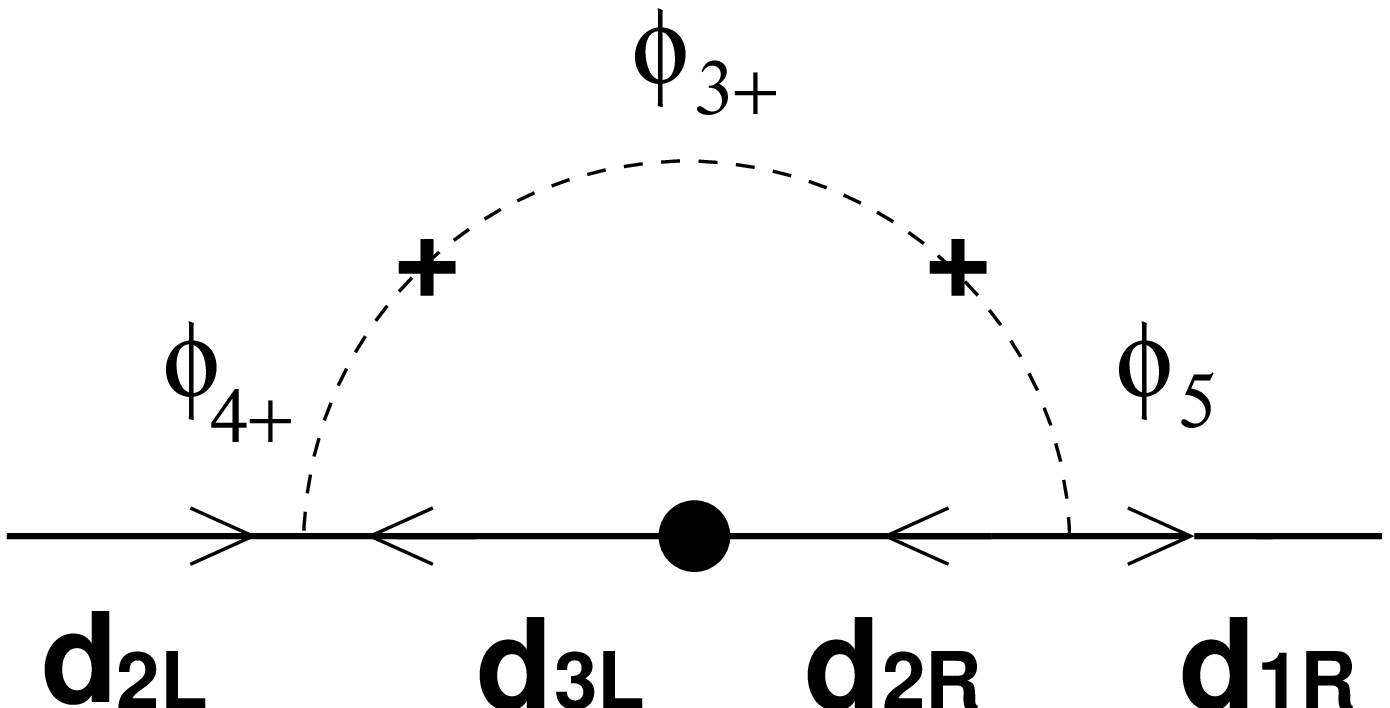} &
\includegraphics[height=1.6cm]{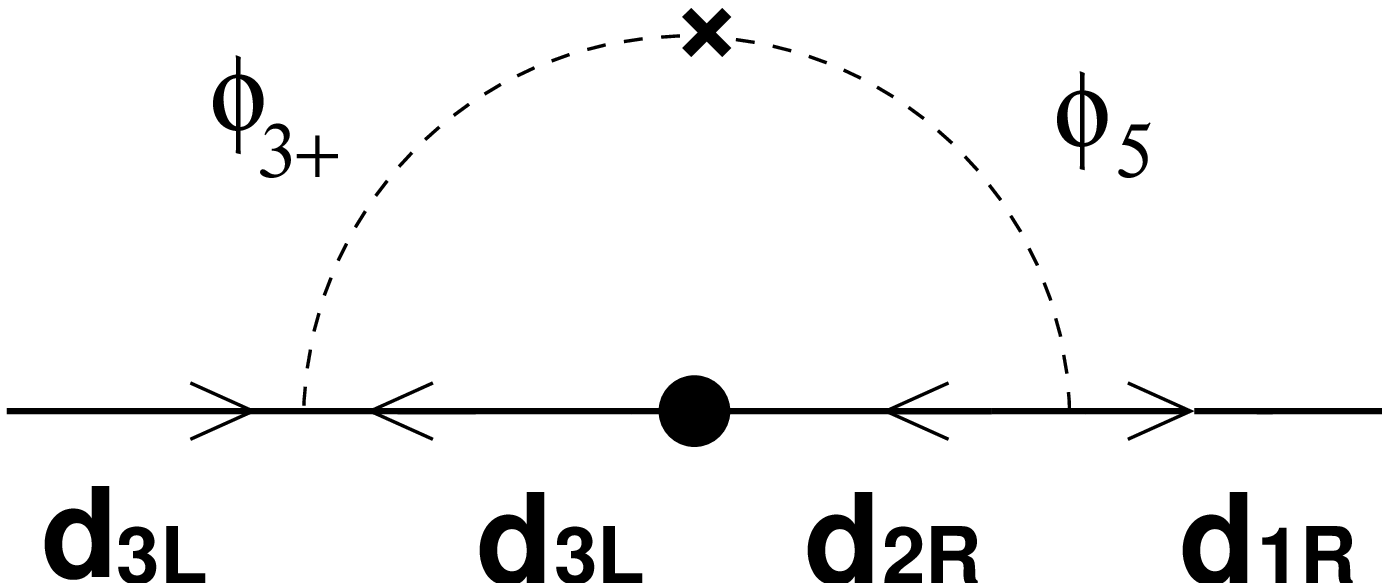}
\\
\includegraphics[height=1.8cm]{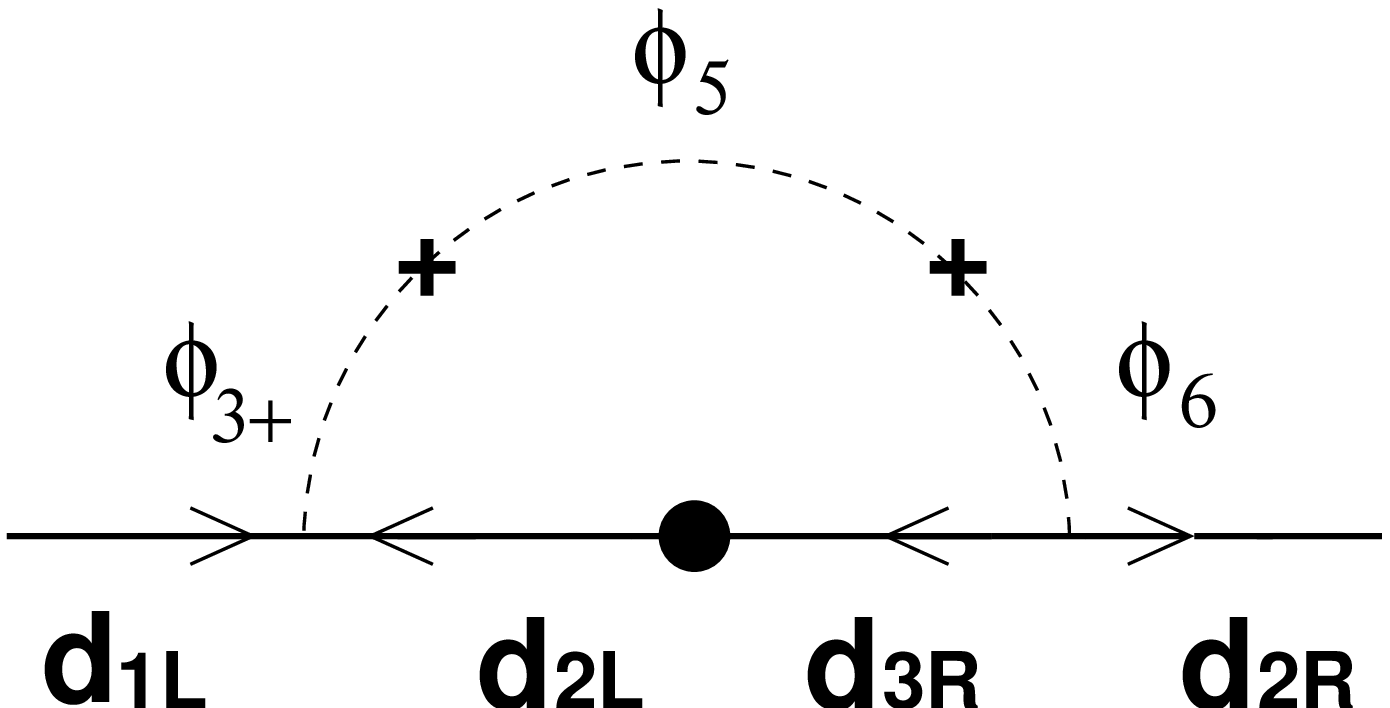} &
\includegraphics[height=1.8cm]{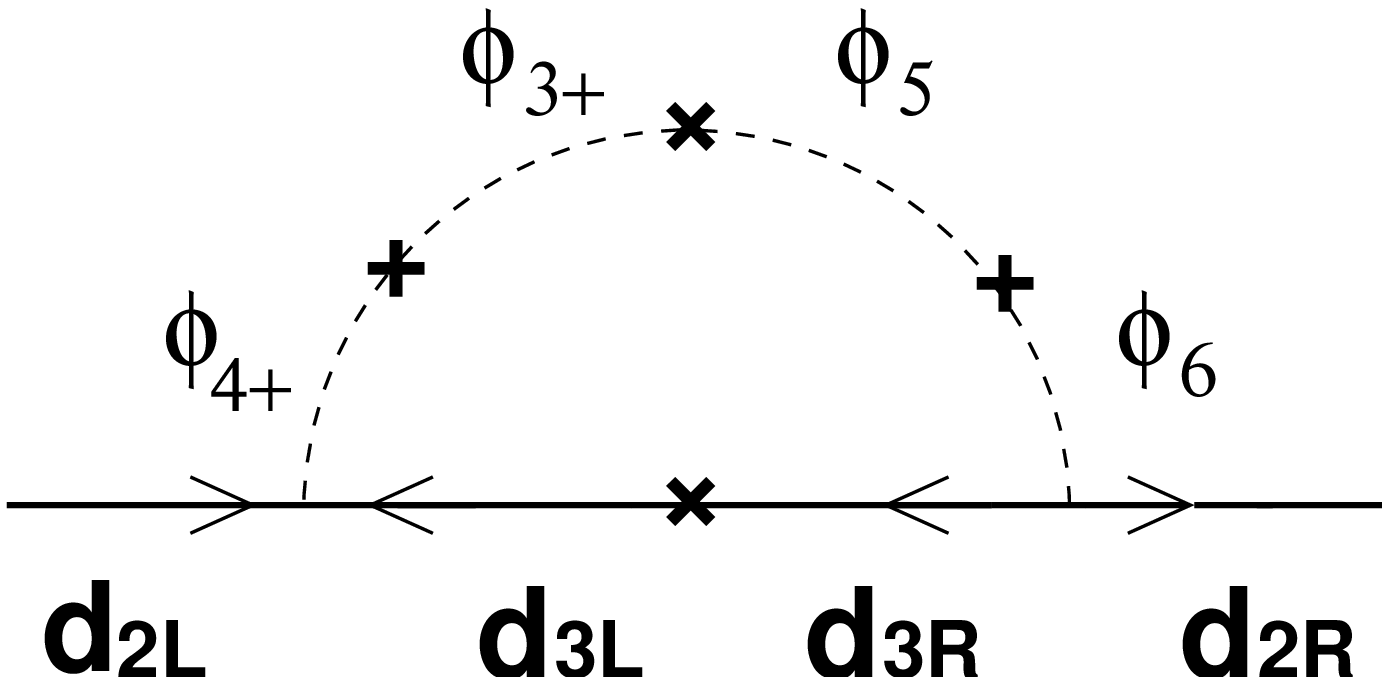} &
\includegraphics[height=1.8cm]{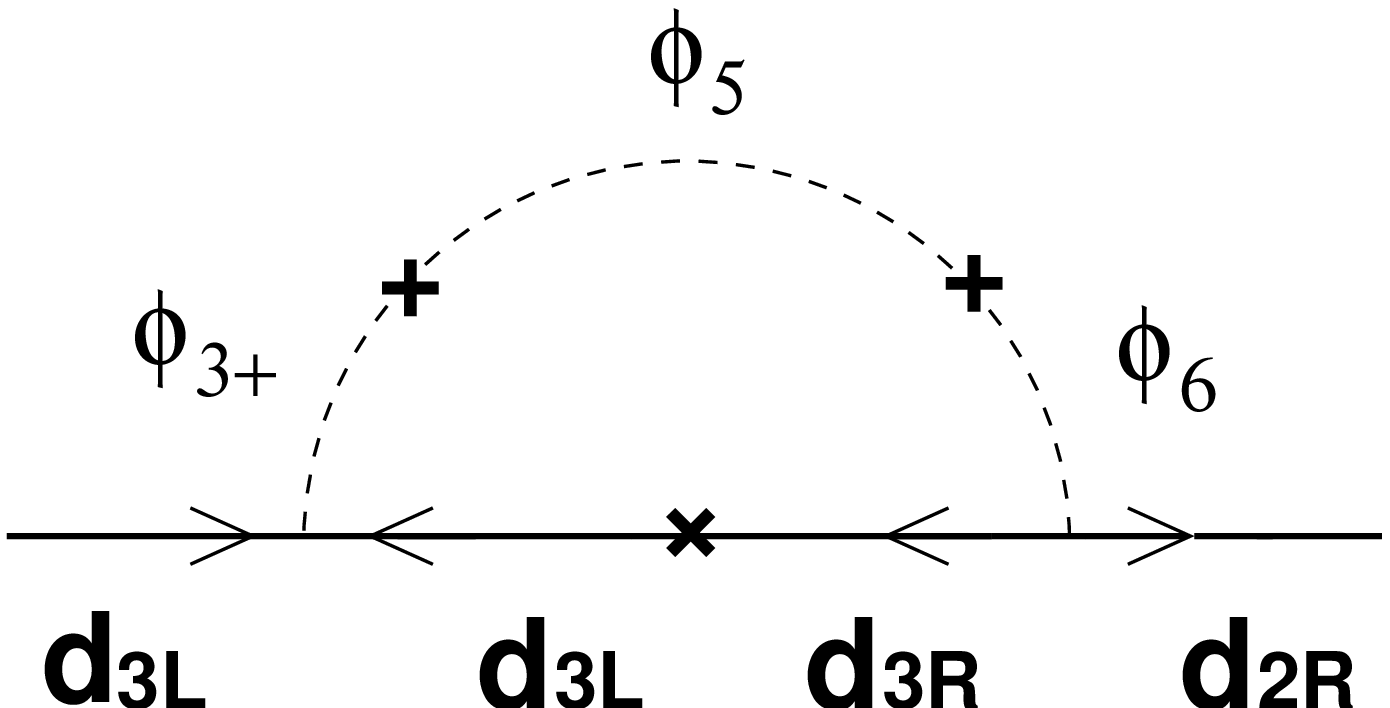} 
\\
\includegraphics[height=1.6cm]{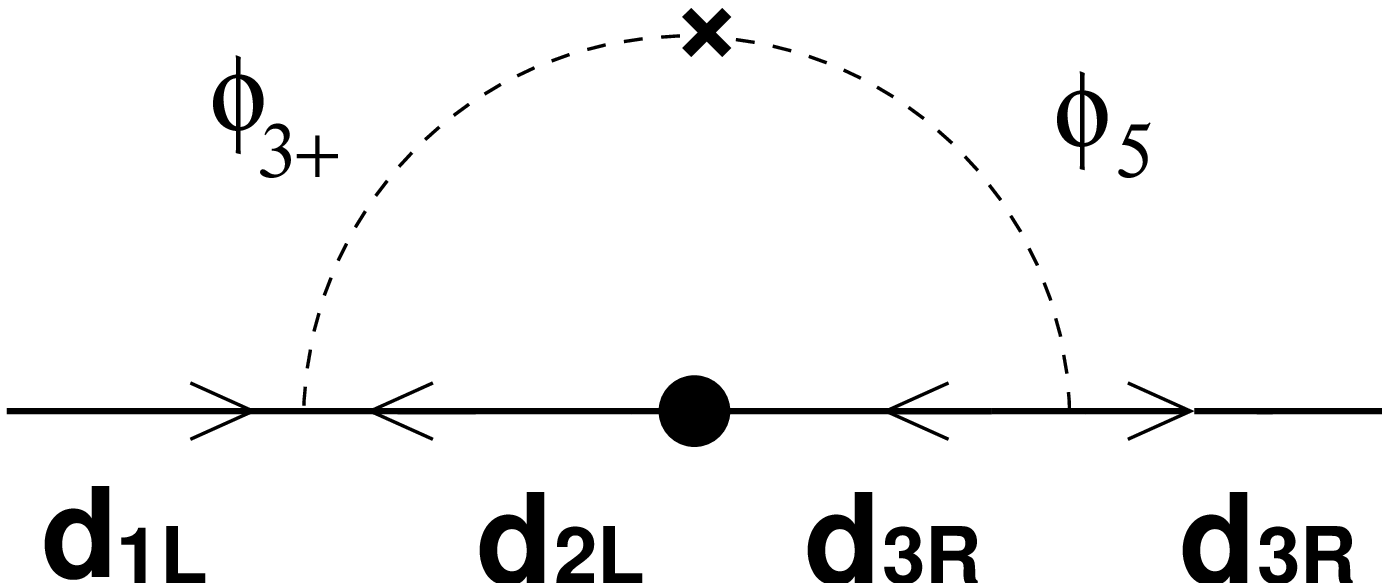} &
\includegraphics[height=1.8cm]{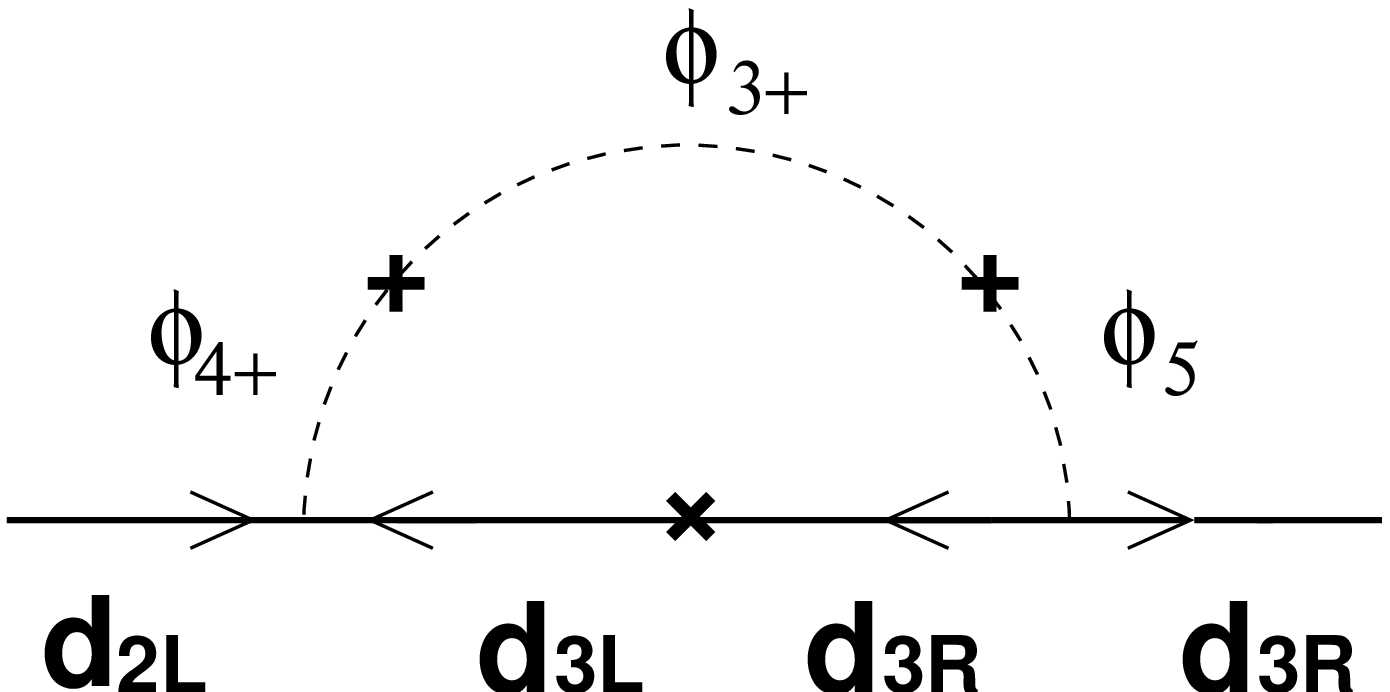} &
\includegraphics[height=0.9cm]{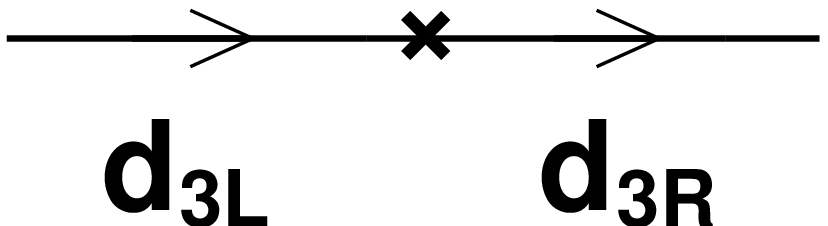}
\end{tabular} 
\caption{\label{mcd} Mass matrix elements for d quarks.}
\end{figure}

\begin{figure}
\begin{center}
\subfigure[]{\centering\includegraphics[height=1.8cm]{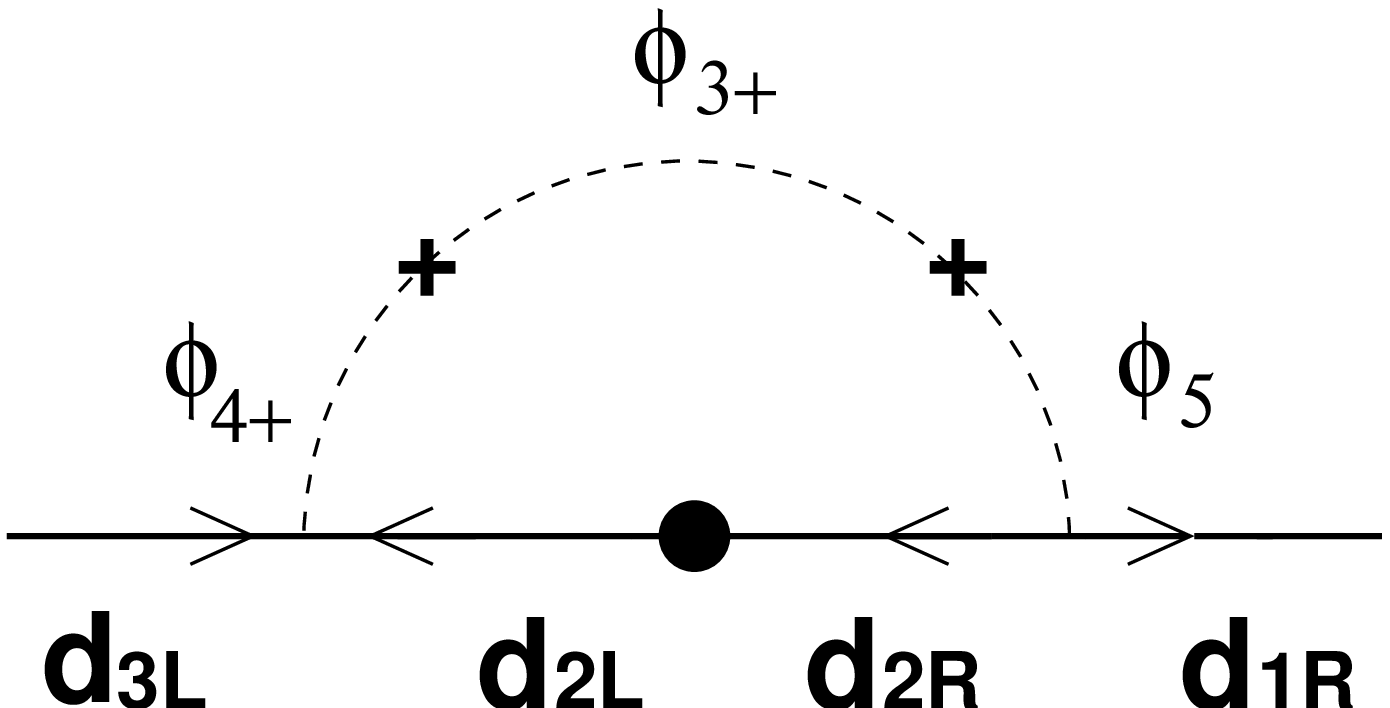}}
\subfigure[]{\centering\includegraphics[height=1.8cm]{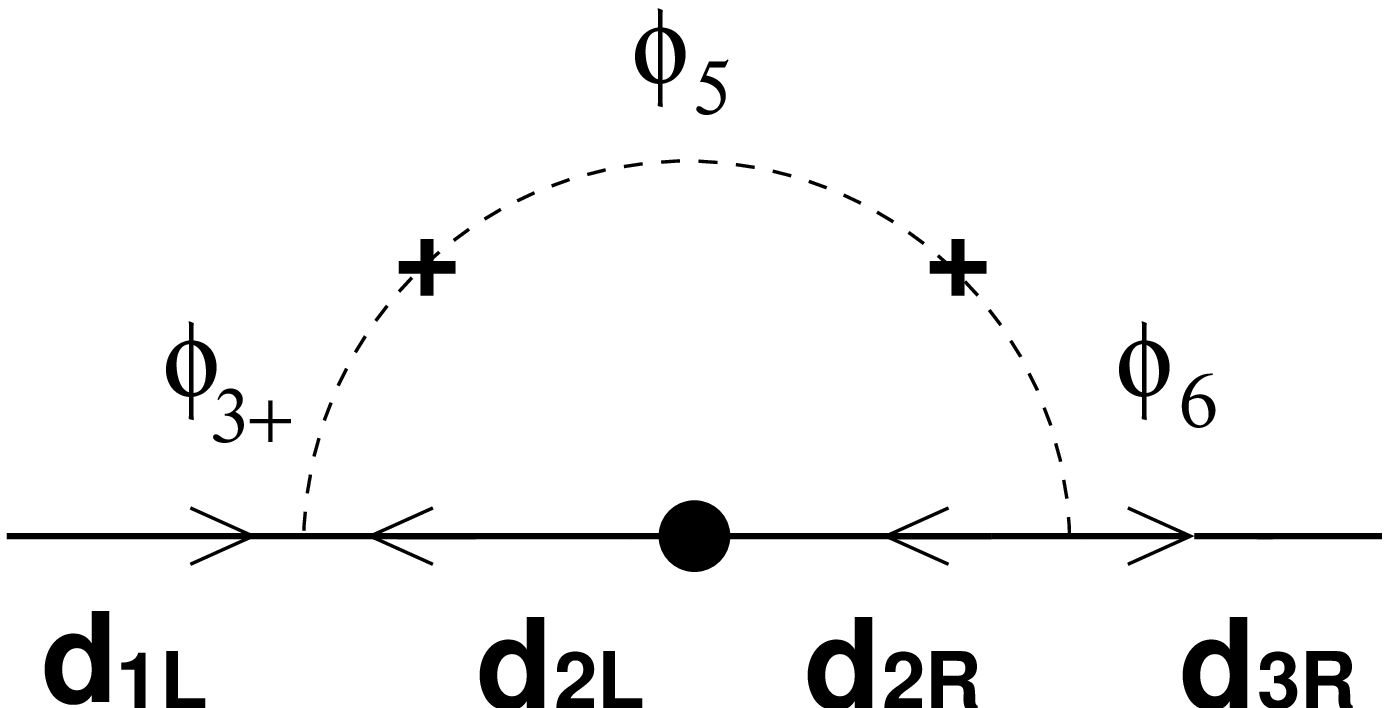}}
\caption{Additional diagrams that contributes to mass matrix elements (a)(1,3)
and (b) (3,1).}
\label{m31b}
\end{center}
\end{figure}

In general there could be mass contributions of  two types as depicted in Fig. \ref{M2fig}. In the present model however, only the Fig. \ref{mcd} diagrams for the charge $-1/3$ quark mass matrix elements and similar ones for the  charge $2/3$  and $-1$ sectors do contribute (these type of diagrams were first introduced in \cite{ma}); in the Fig. \ref{mcd} diagrams the cross means tree level mixing and the black circle means one loop mixing. The diagrams in Fig. \ref{m31b}a and \ref{m31b}b should be added to the matrix elements (1,3) and (3,1), respectively.
%%%%%%%%%%%%%%%%%%%%%%%
%%%%%%%%%%%%%%%%%%%%%%%

In the one loop contribution to the mass matrices for the different
charged fermion sectors only the third family of fermions appears in the
internal lines.  This generates a rank 2 matrix, which once diagonalized
gives the physical states at this approximation. Then using these mass
eigenstates the next order contribution are computed, obtaining a matrix 
of rank 3.  After the diagonalization of this matrix the mass
eigenvalues and eigenstates are obtained (A quark mass mechanism with some similar features to the one proposed here is given in \cite{ma}).

Notice that due to the scalar mixing, in all the loop diagrams of 
Fig. \ref{mcd} and \ref{m31b}, the divergences cancel out in each one of 
these diagrams as is physically expected, giving rise to finite 
contributions to the mass matrices.

Explicitly, the non vanishing contributions from the diagrams of Fig. 
\ref{mcd} to the mass terms $ \bar{d}_{iR} d_{jL} \Sigma^{(1)}_{ij} + h.c.$ read  at one loop
\begin{equation}\label{sigma22} \Sigma_{22}^{(1)} = 3 m_b^{(0)}
\frac{Y_I^2}{16\pi^2} \sum_k U_{1k}U_{4k}f(M_k,m_b^{(0)}),
\end{equation} 
%\begin{equation}\label{sigma23}
%\Sigma_{23}^{(1)}=3 m_b^{(0)}\frac{Y_I^2}{16\pi^2}\sum_kU_{2k}U_{4k}
%f(M_k,m_b^{(0)}),
%\end{equation}
%\begin{equation}\label{sigma32}
%\Sigma_{32}^{(1)} = 3 m_b^{(0)} \frac{Y_I^2}{16\pi^2} \sum_k
%U_{1k}U_{3k}f(M_k,m_b^{(0)}),
%\end{equation}
and similar ones for $\Sigma_{23}^{(1)}$ and $\Sigma_{32}^{(1)}$,
where $m_b^{(0)}$ is the tree level contribution to the b quark mass, the 3 
is a color factor, U is the orthogonal matrix which diagonalizes the mass 
matrix of the charge $2/3$  scalars,  
$(\phi_{4}, \phi_{3}, \phi_5, \phi_6)^T=U(\sigma_1, \sigma_2, \sigma_3, 
\sigma_4)^T,$
where $\sigma_i$ are the eigenfields with eigenvalues $M_i$, and
$$f(a,b) \equiv \frac{1}{a^2-b^2}[a^2ln\frac{a^2}{b^2}],$$
which is just a logarithmic contribution when $a^2 \gg b^2$. The
resulting second rank mass matrix at this level is thus

\begin{equation}\label{mass1}
M_d^{(1)} = \left( \begin{array}{ccc}
 0 & 0 & 0 \\
 0 & \Sigma_{22}^{(1)} & \Sigma_{23}^{(1)} \\
 0 & \Sigma_{32}^{(1)} & m_b^{(0)}
\end{array}  \right).
\end{equation}
At effective two loops we obtain the following expressions:\\
\begin{eqnarray}\label{sigma11.2}
%\begin{equation}\label{sigma11.2}
\Sigma_{11}^{(2)} = 3 \frac{Y_I^2}{16\pi^2} \sum_{k,i}m_i^{(1)}
(V^{(1)}_{dL})_{2i}(V^{(1)}_{dR})_{2i}U_{2k}U_{3k}f(M_k,m_i^{(1)}), \nonumber \\
%\end{equation}  
%
%\begin{equation} \label{sigma12.2}
%\Sigma_{12}^{(2)} = 3 \frac{Y_I^2}{16\pi^2} \sum_{k,i}m_i^{(1)}
%(V^{(1)}_{dL})_{3i}(V^{(1)}_{dR})_{2i}U_{1k}U_{3k}f(M_k,m_i^{(1)}),
%\end{equation}
%\begin{eqnarray}\label{sigma13.2}
\Sigma_{13}^{(2)} = 3 \frac{Y_I^2}{16\pi^2} \sum_{k,i}m_i^{(1)}
(V^{(1)}_{dL})_{3i}(V^{(1)}_{dR})_{2i}U_{2k}U_{3k}f(M_k,m_i^{(1)}) \nonumber \\
+ 3 \frac{Y_I^2}{16\pi^2} \sum_{k,i}m_i^{(1)}
(V^{(1)}_{dL})_{2i}(V^{(1)}_{dR})_{2i}U_{1k}U_{3k}f(M_k,m_i^{(1)}),
\end{eqnarray} 

%\begin{equation}\label{sigma21.2}
%\Sigma_{21}^{(2)} =3 \frac{Y_I^2}{16\pi^2} \sum_{k,i}m_i^{(1)}
%(V^{(1)}_{dL})_{2i}(V^{(1)}_{dR})_{3i}U_{2k}U_{4k}f(M_k,m_i^{(1)}),
%\end{equation} 

%\begin{eqnarray}\label{sigma31.2}
%\Sigma_{31}^{(2)} = 3 \frac{Y_I^2}{16\pi^2} \sum_{k,i}m_i^{(1)}
%(V^{(1)}_{dL})_{2i}(V^{(1)}_{dR})_{3i}U_{2k}U_{3k}f(M_k,m_i^{(1)})
%\\\nonumber
%+
%3 \frac{Y_I^2}{16\pi^2} \sum_{k,i}m_i^{(1)}
%(V^{(1)}_{dL})_{2i}(V^{(1)}_{dR})_{2i}U_{2k}U_{4k}f(M_k,m_i^{(1)}),
%\end{eqnarray} 
and the corresponding ones for $\Sigma_{12}^{(2)}$, $\Sigma_{21}^{(2)}$ 
and $\Sigma_{31}^{(2)}$,
the k (i) index goes from 1 to 4 (from 2 to 3), $V^{(1)}_{dL}$ and 
$V^{(1)}_{dR}$ are the unitary matrices which diagonalize $M_d^{(1)}$ of 
equation (\ref{mass1}) and $m_i^{(1)}$ are the eigenvalues. Therefore at 
two loops the mass matrix for d quarks becomes:

\begin{equation}\label{mass2}
M_d^{(2)} = \left( \begin{array}{ccc}
\Sigma_{11}^{(2)} & \Sigma_{12}^{(2)} & \Sigma_{13}^{(2)} \\
\Sigma_{21}^{(2)} & m_2^{(1)} & 0 \\
\Sigma_{31}^{(2)} & 0 &  m_3^{(1)} 
\end{array} \right).
\end{equation}

For the up quark and charged lepton sectors the procedure to obtain the masses is completely
analogous.
% That is, the mass terms for the up sector come from graphs like those in Fig. \ref{mcd} and \ref{m31b}, but replacing the  $\phi_{4}$,$\phi_{3}$, $\phi_5$ and $\phi_6$ scalar fields by $\phi_{4}$,$\phi_{3}$, $\phi_7$ and $\phi_8$ and the quarks $d_i$ by the quarks $u_i$.

The CKM matrix takes the form
\begin{equation}\label{ckmmatrix}
V_{CKM} = (V_{uL}^{(2)} V_{uL}^{(1)})^{\dag} V_{dL}^{(2)} V_{dL}^{(1)},
\end{equation}
where the unitary matrices $V_{uL}^{(1)}$ and $V_{uR}^{(1)}$ diagonalize
$M_u^{(1)}$, and $V_{uL}^{(2)}$ and $V_{uR}^{(2)}$ diagonalize
$M_u^{(2)}$, with an analogous notation used for the down sector.

It is important to mention here that the textures, particularly the
zeros in the scalar and fermion mass matrices
(Eqs. \ref{scalarmass} and \ref{mass2}) are neither accidental nor
imposed; they are just a direct consequence of
the mass mechanism and of the gauge symmetry of the model.

\subsection{Rare Decays and Anomalous Magnetic Moments}

\begin{figure}[!ht]
%\subfigure{
\centering\includegraphics[height=3cm]{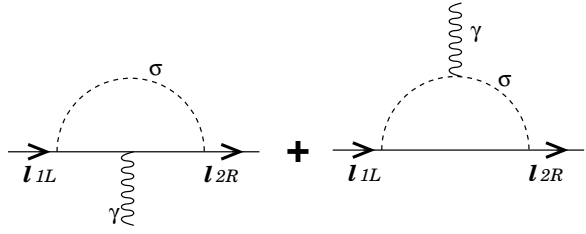}
%}
%\hspace{0.5cm}+\hspace{0.5cm}
%\subfigure{\includegraphics[height=3cm]{DiagD2.eps}}
\caption{\label{DiagDecay} Diagrams for the process $l_1 \rightarrow 
l_2\gamma$ where a scalar mass eigenstate $\sigma$ is involved.}
\end{figure}

Considering two equally charged fermions the  process 
$f_1\rightarrow f_2 \gamma$ is written as \cite{BLee} 
\begin{equation}
iM(l_{1}(p_1) \rightarrow l_{1}(p_2) + \gamma) = 
 i\bar{u}_{2}(p_2) \left( \epsilon^{\mu} \gamma_{\mu} F_{1}^{V}(0) \delta_{f_1 f_2} + 
\frac{\sigma _{\mu \nu} q^{\nu} \epsilon ^{\mu}}{m_1 + m_2} (F_{2}^{V}(0) + F_{2}^{A}(0) \gamma _5) \right) u_{1} (p_1)
\label{amplitud}
\end{equation}

where $F_2^{A (V)}$  gives the anomalous magnetic moment (electric dipole 
moment) for the fermion $f_1$ when $f_1 = f_2$.
The diagrams for the processes are shown in figure \ref{DiagDecay}, in 
these diagrams $\sigma$ stands for a mass eigenstate scalar field.

The respective evaluation of these diagrams gives
%%%%%%%%%%%%%%%%%%%%%%%%%%%%%
\begin{eqnarray}
iA_1 = \frac{Y_{l}^{2} q_{e}}{16 \pi^{2}} N(m_{\sigma}, m_{i}) \bar{e}_{2R} 
i \sigma^{\mu \nu} q_{\nu} \epsilon_{\mu} e_{1L} & \mbox{and}&
iA_2 =\frac{Y_{l}^{2} q_{e}}{16 \pi^{2}}  N(m_{\sigma}, m_{i}) \bar{e}_{2L} 
i \sigma^{\mu \nu} q_{\nu} \epsilon_{\mu} e_{1R}
\label{Amp_Decay}
\end{eqnarray}
where  $N(m_{\sigma},m_{i}) = \frac{m_{i}}{m_{\sigma}^{2}} \left[ln \frac{m_{\sigma}^{2}}{m_{i}^{2}} - \frac{1}{2}\right]$, the relation $m_{\sigma} \gg m_i$ is assumed to be held and the second amplitude comes from the diagrams where $L$ and $R$ are interchanged.
Notice that due to scalar field mixing the contribution of these loops are finite as those in the mass case.

Due to the fermion mixing  matrices structure (see the numerical results) 
the diagrams that make the largest contribution to the AMM of the leptons 
are; for the electron, the diagram with the muon inside the loop, and for the 
muon and tau, the diagrams with the tau as the inner fermion.
The expression for the scalar contribution to the muon AMM is
\begin{eqnarray}
a_{\mu} &=& \frac{m_{\mu}Y_{l}^{2}}{16 \pi ^{2}} (V_{eL}^{l})_{22}(V_{eR}^{l})_{22}(G^{\mu_{R}} + G^{\mu_{L}}) %\nonumber \\
%& \approx & \frac{m_{\mu}Y_{l}^{2}}{16 \pi ^{2}} (G^{\mu_{R}} +
%G^{\mu_{L}})
\label{amm-mu}
\end{eqnarray}
where
\begin{eqnarray}
G^{\mu R}= \sum_{k,i} U_{1k}^{l} U_{4k}^{l}(V_{eL}^{l})_{3i}(V_{eR}^{l+})_{i3} 
N(m_{\sigma_{k}},m_{i}) \nonumber \\
G^{\mu L} = \sum_{k,i} U_{4k}^{l} U_{1k}^{l}(V_{eR}^{l})_{3i}(V_{eL}^{l+})_{i3} N(m_{\sigma_{k}},m_{i})
\label{amm-mu2}
\end{eqnarray}
and analogous expression (with a suitable indices change) are held for the 
$e$ and  $\tau$ leptons.

For the branching ratios of flavor changing decays the following 
expressions are obtained

\begin{multline}
\Gamma (\mu \rightarrow e + \gamma) = 
(m_{\mu} + m_{e})^{2} 
\frac{m_{\mu} Y_{l}^{4}}{(16)^{3} \pi ^{5}} 
\left(1 - \frac{m_{e}}{m_{\mu}}\right)^{2} 
\left(1-\frac{m_{e}^{2}}{m_{\mu}^{2}}\right) \\ 
\left( |(V_{eL}^{l})_{22}(V_{eR}^{l})_{11}G^{\mu_{L}e_{R}}|^{2} + 
|(V_{eL}^{l})_{11} (V_{eR}^{l})_{22}G^{\mu_{R}e_{L}}|^{2} \right) \\
\approx \frac{m_{\mu}^{3} Y_{l}^{4}}{(16)^{3} \pi ^{5}} 
\left(|G^{\mu_{L}e_{R}}|^{2} + |G^{\mu_{R}e_{L}}|^{2} \right) 
\label{Branching1}
\end{multline}
with 
\begin{eqnarray}
G^{\mu_{R}e_{L}} = \sum_{k,i} U_{1k}^{l} U_{3k}^{l} (V_{eR}^{l})_{2i} (V_{eL}^{l+})_{i3} N(m_{\sigma_{k}},m_{i}) \nonumber \\
G^{\mu_{L}e_{R}} = \sum_{k,i} U_{4k}^{l} U_{2k}^{l} (V_{eL}^{l})_{2i} (V_{eR}^{l+})_{i3} N(m_{\sigma_{k}},m_{i})
\label{Branching2}
\end{eqnarray}
and analogous expressions for $\tau \rightarrow \mu \gamma$ and $\tau 
\rightarrow  e \gamma$.

\begin{figure}[h]
\begin{center}
%\subfigure[Generic tree level Flavor Changing process]{
\includegraphics[height=3cm]{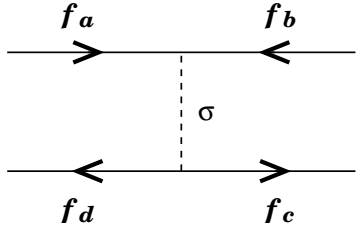}%}
%\subfigure[Generic one loop level Flavor Changing process]{\includegraphics[height=3cm]{Drare2.eps}}
\end{center}
\caption{\label{Drare}Generic tree level flavor changing processes. In these type 
of diagrams $\sigma$ stands for a scalar 
mass eigenstate.}
\end{figure}

\begin{figure}
\begin{center}
\subfigure{\includegraphics[height=3cm]{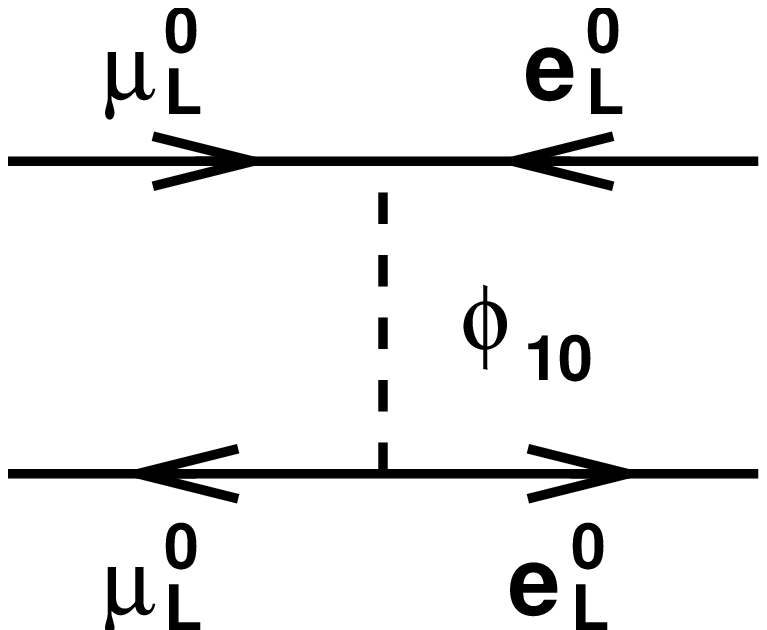}}
\subfigure{\includegraphics[height=3cm]{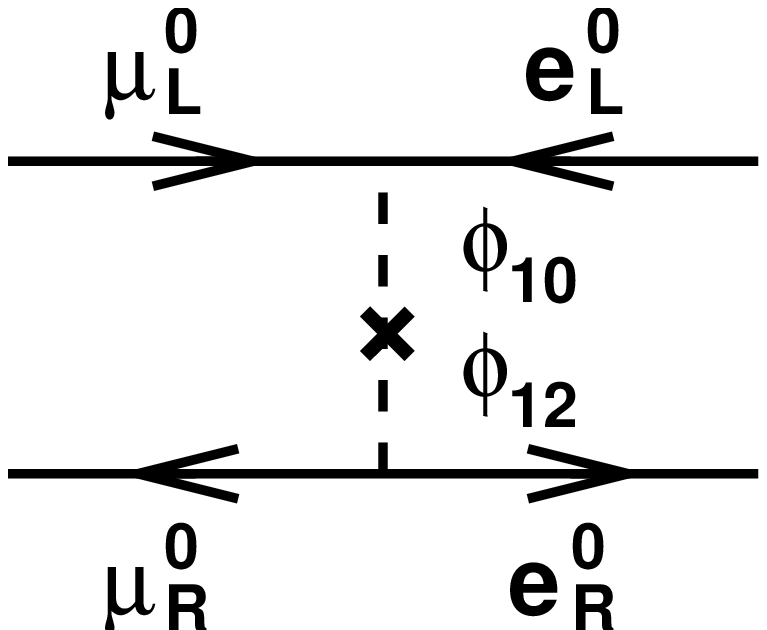}}
\subfigure{\includegraphics[height=3cm]{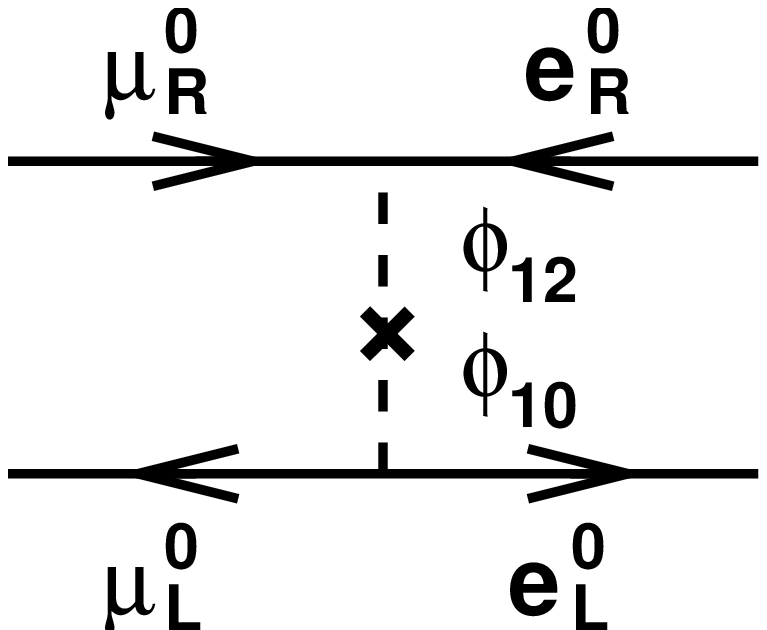}}
\subfigure{\includegraphics[height=3cm]{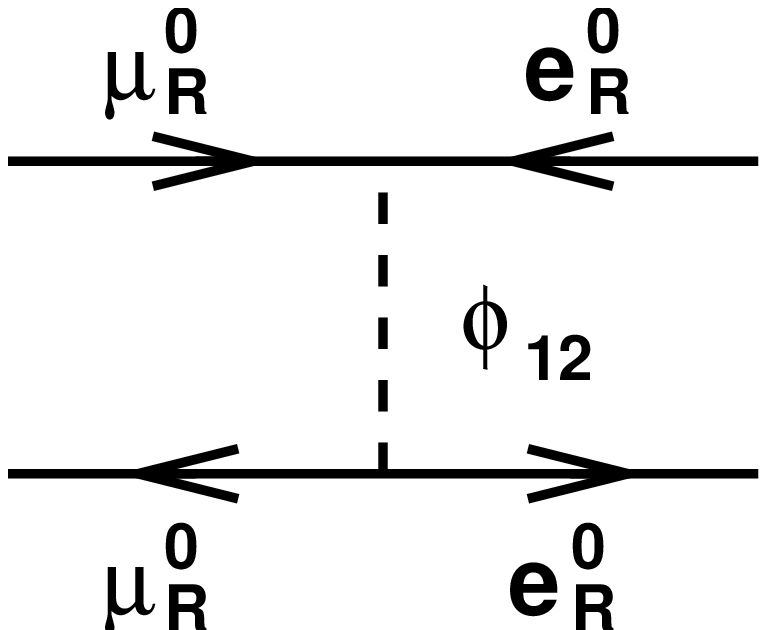}}
\end{center}
\caption{\label{FCDiag} Diagrams contributing to muon flavor 
changing decays.}
\end{figure}

The scalar fields ($\phi_9, \phi_{10}, \phi_{12}, \phi_{11}$) allow tree 
level flavor changing decays due to the  mixing among themselves, e.g. 
Fig. \ref{Drare}. In particular, the processes which could be compared 
with experimental bounds are $\tau \rightarrow  3\mu$, $\tau 
\rightarrow 2\mu e$, $\tau \rightarrow 2 e \mu$, $\tau \rightarrow 3 e$, 
$\mu \rightarrow 3 e $, with diagrams like those in Fig. \ref{FCDiag}. For 
these processes the available phase space should be evaluated, and it is well known that  
for one particle of mass $M$ and momentum $P$  
decaying to three particles with momenta $p_1$, $p_2$ and $p_3$ and masses 
$m_1, m_2$ and $m_3$  \cite{spearman} the branching ratio is given by

\begin{multline} 
\Gamma (P \rightarrow p_1 p_2 p_3) = \frac{2 m_1 m_2
m_3}{(2 \pi)^3 n_{f}!} \int dE_1 dE_2 \Theta (E_1,E_2) |M|^2, \\ \mbox{
which in this case reads}\\ = \frac{\alpha^{2} Y_{l}^4}{4(2 \pi)^3 n_{f}!}
\left| \sum_k \frac{\mathfrak N(U^{l}(k))}{m_{\sigma_k}^{2}} \right|^2
\int dE_1 dE_2 \Theta(E_1,E_2) \\ E_1(m^2 + m_1^2 -m_2^2 -m_3^2 -2mE_1).
\label{de-c11} 
\end{multline}

$\Theta(E_1,E_2)$ represents the available phase space for the process,
$\alpha$ and ${\mathfrak N}(U^{l}(k))$ are shown in Table
\ref{tabladealpha}. The fermion mixing matrix structure was taken into
account and only the largest contributions are evaluated.
From the numerical fit the lepton mixing matrices have the form
\begin{eqnarray}
V = & V_{eR}^{l} \approx V_{eL}^{l} \approx & 
\left(
\begin{array}{ccc}
1 & \lambda & \lambda \\
\lambda & 1 & \lambda^2 \\
\lambda & \lambda^2 & 1 \\
\end{array}
\right),
\end{eqnarray}
with $\lambda \sim 10^{-4}$.

\begin{table}[h] 
\begin{center} 
\begin{tabular}{lcr} 
\toprule 
Process & $\alpha$ & ${\mathfrak N}(U^{l}(k))$ \\\hline %\midrule 
$\tau \rightarrow \mu \mu
\mu$ & $2 \lambda$ & $(U_{1k}^{l}+U_{4k}^{l})(U_{2k}^{l}+U_{3k}^{l})$ \\
$\tau \rightarrow \mu \mu e$ & $2$ &
$(U_{1k}^{l}+U_{4k}^{l})(U_{2k}^{l}+U_{3k}^{l})$ \\ $\tau \rightarrow e e
\mu $ & $2 \lambda$ & $(U_{2k}^{l} + U_{3k}^{l})(U_{1k}^{l}+U_{4k}^{l}+1)
+ 2 U_{2k}^{l} U_{3k}^{l}$ \\ $\tau \rightarrow e e e $ & $2 \lambda ^2$ &
$(U_{2k}^{l} + U_{3k}^{l})(U_{1k}^{l}+U_{4k}^{l}+1) + 2 U_{2k}^{l}
U_{3k}^{l}$ \\ $\mu \rightarrow eee$ & $\lambda$ & $U_{2k}^{l} +
U_{2k}^{l} U_{3k}^{l} + U_{3k}^{l}$ \\ 
\bottomrule 
\end{tabular} 
\caption{$\alpha$ and ${\mathfrak N}(U^{l}(k))$ for tree level flavor changing
decays.} 
\label{tabladealpha} 
\end{center} 
\end{table}

\section{Numerical Evaluation} 

\subsection{Experimental Values} 

Since quarks are confined inside hadrons, their masses cannot  be directly
measured. So, the quark mass parameters in the SM Lagrangian depend both
on the renormalization scheme adopted to define the theory and on the
scale parameter $\mu$ where the theory is being tested. In the limit where
all quark masses are zero, the SM has an $SU(3)_L\otimes SU(3)_R$ chiral
symmetry which is broken at a scale $\Lambda_\chi\simeq 1$GeV. To
determine the quark mass values SM perturbation theory must be used at an
energy scale $\mu>>\Lambda_\chi$ where non perturbative effects are
negligible.

For illustration, the allowed ranges of quark masses\cite{pdg} in the modified 
minimal subtraction scheme $(\overline{MS})$ are\cite{nari}:

\begin{eqnarray}\nonumber
m_u(1.GeV)&=&2-6.8\; MeV.\\ \nonumber 
m_d(1.GeV)&=&4-12\; MeV.\\ \nonumber
m_s(1.GeV)&=&81-230\; MeV.\\  \nonumber
m_c(m_c)&=&1.1-1.4\; GeV.\\ \nonumber
m_b(m_b)&=&4.1-4.4\; GeV.\\ \nonumber
m_t(Exp)&=&173.8\pm 5.2 GeV 
\end{eqnarray}

To find the relative magnitude of different quark masses in a meaningful
way, one has to describe all quark masses in the same scheme and at the
same scale. In this analysis the quark masses are calculated at an
energy scale $\mu_m$ such that $M_Z<\mu_m<M_X\simeq v_2$, where $M_X$ is
the mass scale and where $U(1)_X$ is spontaneously broken. Since in this model
there is no mixing between the Standard Model Z boson and its horizontal
counterpart, $v_2$ could be as low as the electroweak breaking scale.
For simplicity, assume that these calculations are meaningful at the
electroweak breaking scale and from the former values for the quark masses
 calculate, in the $\overline{MS}$ scheme, the quark masses at the
$m_Z$ scale\cite{fritzsch}. Those values calculated in the cited 
reference are presented in Table \ref{masas:fritzsch}.

On the other hand, the CKM matrix elements are not ill defined and they
can be directly measured from the charged weak current in the SM. For
simplicity the assumption was made that they are real, and as discussed in
Ref.\cite{fusaoka}, they are almost constant in the interval $M_Z<\mu<$ a
few TeV. Their current experimental value \cite{pdg} and the estimated ones 
are given in Table \ref{fritzsch:res}.

\subsection{Evaluation of the Parameters}

In order to test the model using the least possible number of free
parameters, let the scalar mass matrices be written in the following 
form: 

%%%%%%%%%
\begin{eqnarray}\label{emm}
M_{2/3}^2 = \left(
\begin{array}{cccc}
a_+ & b & 0        & 0 \\
b & a_+ & c_+  & 0 \\
0 & c_+ & a_+  & d_+ \\
0 & 0 & d_+ & a_+
\end{array} \right)
& \mbox{and}\hspace{1cm} &
M_{4/3}^2 = \left(
\begin{array}{cccc}
a_- & b & 0        & 0 \\
b & a_- & c_-  & 0 \\
0 & c_- & a_-  & d_- \\ 
0 & 0 & d_- & a_-
\end{array} \right).
\end{eqnarray} \\

Using the central value of the CKM elements in the PDG book\cite{pdg} and
the central values of the six quark masses at the Z mass scale\cite{fritzsch},
 the $\chi^2$ function is built in the  parameter space.
 Diagonalization of involved matrices is achieved using LAPACK
\cite{lapack} routines,
 and the $\chi^2$ function is minimized using MINUIT from the CERNLIB
packages\cite{minuit}; both Monte Carlo and standard routines were used in
the minimization process. The tree level masses of the top, bottom and 
tau were restricted to be around the central
values $\pm$ 10 \% in order to assure consistency with the assumption that
radiative corrections are small. The numerical values for the parameters
are shown in Table \ref{parametros:f}.  With the numeric values which minimize 
$\chi^2$ numerical values for $m_q(m_t)$ with
$q=u,d,c,s,t,b$, and $(CKM)_{ij}$ with $i,j=1,2,3$ are shown in Table
\ref{masas:fritzsch}.

For the sake of comparison, the same calculations but now using the
central values of the six quark masses at the $M_t$ scale \cite{fusaoka} 
are repeated. 
The numerical results are presented in Table 
\ref{fusaoka:res} and
\ref{parametros:fu}.% and \ref{fusaoka:res}.

\begin{table}[!ht] 
\centering 
\subtable[\label{masas:fritzsch}Quark Masses (MeV)]{%
\begin{tabular}{cccc} \toprule
 &\em Central Value &\em Range &\em Result \\\hline %\midrule 
$m_d$ &3.55 &
1.8 a 5.3 & 3.72 \\ $m_s$ &67.5 &30 a 100 & 50.7 \\ $m_b$ &2900 & 2800 a
3000 & 2470 \\ $m_u$ &1.9 &0.9 a 2.9 & 1.83 \\ $m_c$ &605 & 530 a 680 &
224.9 \\ $m_t$ &174000 &168000 a 180000 & 176400 \\\bottomrule
\end{tabular}} \hspace{5mm}
\subtable[CKM Matrix elements]{%
\begin{tabular}{cccc}\toprule &\em
Central Value &\em Range &\em Results \\\hline %\midrule 
$CKM_{11}$ & 
0.97495
&0.9742 a 0.9757 & 0.9747 \\ $CKM_{12}$ & 0.22250 &0.219 a 0.226 & 0.2235 
\\ $CKM_{13}$ & 0.00350 &0.002 a 0.005& 0.0033 \\ $CKM_{21}$ & 0.22200
&0.219 a 0.225& 0.2234 \\ $CKM_{22}$ & 0.97405 &0.9734 a 0.9749& 0.9737 \\
$CKM_{23}$ & 0.04000 &0.037 a 0.043& 0.0434 \\ $CKM_{31}$ & 0.00900 &0.004
a 0.014& 0.0064 \\ $CKM_{32}$ & 0.03900 &0.035 a 0.043& 0.0439 \\
$CKM_{33}$ & 0.99915 &0.9990 a 0.9993& 0.9990 \\ \bottomrule
\end{tabular}} 
\caption{\label{fritzsch:res}Fit results using the mass values at Z 
mass scale.}
\end{table}

\begin{table}[!ht] 
\centering
\begin{tabular}{cc|cc}\toprule 
\em Parameter &\em Value ($\times 10^{6}
TeV^2$)&Parameter & Value (MeV) \\\hline %\midrule
 $a_{+}$ & 19.6111 & M0b &  2474.12  \\
 $b$ &      0.105233 & M0t & 176061.8  \\\cmidrule{3-4}
 $c_{+}$ & 19.3262 &$Y_{Q}$ &  16.95 \\
 $d_{+}$ & 0.656368 &&\\
 $c_{-}$ & 508.630 && \\
 $d_{-}$ & 6.40954 &&\\
 $a_{-}$ & 9999.19 &&\\\bottomrule \end{tabular}
\caption{\label{parametros:f}Fit parameter values using masses at Z mass 
scale.}
\end{table}

\begin{table}[!ht]
\centering
\subtable[\label{masas:fusaoka}Quark masses ]{%
\begin{tabular}{cccc} \toprule
 &\em Central Value &\em Range &\em Result \\\hline %\midrule 
$m_d$ & 4.49 MeV &  
3.89 a 5.09 MeV & 4.87 MeV \\ $m_s$ &89.4 MeV  &76.9 a 100.9 MeV  & 80.8 
MeV
\\ $m_b$ &2.85 GeV & 2.74 a 2.96 GeV & 2.84 GeV \\ $m_u$ &2.23 MeV& 1.83 a
2.63 MeV & 2.33 MeV \\ $m_c$ &646 MeV& 595 a 700 MeV & 632 MeV \\ $m_t$ 
&171 GeV &169 a 183 GeV & 171.9 GeV \\\bottomrule 
\end{tabular}} \hspace{5mm}
\subtable[CKM matrix elements]{% 
\begin{tabular}{cccc}\toprule &\em Central Value &\em
Range &\em Result \\\hline %\midrule 
$CKM_{11}$ & 0.97495 &0.9742 a 0.9757
& 0.9741 \\ $CKM_{12}$ & 0.22250 &0.219 a 0.226 & 0.2254 \\ $CKM_{13}$ &
0.00350 &0.002 a 0.005& 0.0175 \\ $CKM_{21}$ & 0.22200 &0.219 a 0.225&  
0.2253 \\ $CKM_{22}$ & 0.97405 &0.9734 a 0.9749& 0.9742 \\ $CKM_{23}$ & 
0.04000 &0.037 a 0.043& 0.0036 \\ $CKM_{31}$ & 0.00900 &0.004 a 0.014&   
0.0179 \\ $CKM_{32}$ & 0.03900 &0.035 a 0.043& 0.0003 \\ $CKM_{33}$ &
0.99915 &0.9990 a 0.9993& 0.9998 \\ \bottomrule 
\end{tabular}}
\caption{\label{fusaoka:res}Fit results using masses at top mass scale.} 
\end{table}

\begin{table}[!ht] \centering \begin{tabular}{cc|cc}\toprule \em Parameter
&\em Value ($\times 10^{6} TeV^2$)&Parameter & Value (MeV) 
\\\hline %\midrule   
$a_+$ & 1602.24 & M0b & 2831.64 \\
 b & 0.004034 &M0t& 171950.09 \\\cmidrule{3-4} $c_+$ & 125.391 &$Y_Q$&
14.57 \\ $d_+$ & 51.4477 &&\\ $c_-$ & 1.360099 $\times 10^{-7}$&& \\ $d_-$
& 1.31220&&\\ $a_-$ & 717.0733 &&\\\bottomrule \end{tabular}
\caption{\label{parametros:fu}Fit parameter values using masses at top 
mass scale.}
\end{table}

The results for the charged leptons case using the mass data by 
\cite{fusaoka} are shown in Tables \ref{maslep:fus} and 
\ref{paralep:fus}.

\begin{table}[!ht] \centering \begin{tabular}{ccc}\toprule & Reported Mass
(MeV)& Evaluated Mass (MeV) \\\hline %\midrule 
$M_e$&0.48684727$\pm$0.00000014 &
0.486847282\\ $M_{\mu}$&102.75138$\pm$0.00033 &102.751363 \\
$M_{\tau}$&1746.7$\pm$0.3 & 1746.96\\ \bottomrule \end{tabular}
\caption{\label{maslep:fus}Charged lepton masses at Z mass scale.} 
\end{table}

\begin{table}[!ht] \centering \begin{tabular}{cc|cc}\toprule Parameter &
Value($\times 10^{4}$ $TeV^2$)&Parameter & Value (MeV)\\\hline %\midrule 
a &
52.633013 &$M_{\tau}^{0}$ & 1746.87 \\ b & 50.011234 & & \\ c & 1.8805682
& & \\\cmidrule{3-4} d & 39.978871 &$Y_{L}$ & 3.4124\\ \\\bottomrule
\end{tabular} \caption{\label{paralep:fus}Fit parameter values using 
lepton masses at Z mass scale.} 
\end{table}

The apparently impressive correspondence between the estimated lepton 
masses in this model and the Fusaoka (et. al) calculation is due to the 
large number of free parameters, which in the numerical fit are not as 
constrained as in the quarks case.

Two more comments should be made: First, the values for the parameters 
in the square mass matrices of scalar fields are of order $10^{17} 
(\mbox{MeV})^2$ (see Table
\ref{parametros:f}), so, the scalar physical masses are of order $10^{3}$ 
TeV.
Second, the rounding errors allow us to take safely up to five significative
figures in the masses and in the CKM matrix elements.

As can be seen from Tables \ref{fritzsch:res} and \ref{fusaoka:res}, even 
under the assumption that the CKM matrix elements are real, the numerical 
values are in
good agreement with the allowed experimental results.

%%%%%%%%%%%%%%%%%%%%%%%%
%%%%%%%%%%%%%%%%%%%%%%%%
%%%%%%%%%%%%%%%%%%%%%%%%

\subsection{Anomalous Magnetic Moments and Rare Decays Evaluation}

Using the parameter values from the mass fit, the numerical evaluation  
of scalar contributions to magnetic
moments of the charged leptons and the branching ratio of the rare
decays were made, and they are shown in Tables 
\ref{Tmom} and \ref{tab-res-1}, respectively.

\begin{table}
\begin{center}  
\begin{tabular}{c c}\toprule
\multicolumn{2}{c}{Magnetic Moments}\\\hline%\\\midrule
$a_e$ & $1.3740 \times 10 ^{-17}$  \\
$a_{\mu}$ & $9.1012 \times 10^{-12}$ \\
$a_{\tau}$ & $2.3951 \times 10 ^{-12}$  \\
\bottomrule
\end{tabular}
\caption{Numerical evaluation of scalar contributions to  Anomalous 
Magnetic Moments.
\label{Tmom}}
\end{center}
\end{table}

\begin{table}[h] 
\begin{center} 
\begin{tabular}{c c c} \toprule
\multicolumn{3}{c}{Rare decays} \\\hline %\midrule 
$B(\mu \rightarrow e +
\gamma)$ & $1.3657 \times 10^{-12}$ & $1.2 \times 10^{-11}$ \\ $B(\tau
\rightarrow \mu + \gamma)$ & $2.1440 \times 10^{-12}$ & $1.1 \times
10^{-6}$ \\ $B(\tau \rightarrow e + \gamma)$ & $1.8240 \times 10^{-17}$ &
$2.7 \times 10^{-6}$ \\ \midrule 
$B(\tau \rightarrow \mu \mu \mu)$ & 
$4.1501
\times 10^{-43}$ & $1.9 \times 10 ^{-6}$ \\ $B(\tau \rightarrow \mu \mu
e)$ & $1.3631 \times 10^{-32}$ & $1.5 \times 10^{-6}$ \\ $B(\tau
\rightarrow ee \mu)$ & $1.3698 \times 10^{-15}$ & $1.9 \times 10^{-6}$ \\
$B(\tau \rightarrow eee)$ & $1.9979 \times 10^{-21}$ & $2.9 \times
10^{-6}$\\ $B(\mu \rightarrow eee)$ & $7.0532 \times 10^{-20}$ & $1.0
\times 10^{-12}$ \\\bottomrule
\end{tabular} 
\caption{Branching ratios numerical evaluation. The first column lists the 
process, the second is the evaluation within the $U(1)_{X}$ model, the 
last column shows the experimental bounds. \label{tab-res-1}} 
\end{center} 
\end{table}

\section{CONCLUSIONS}
By introducing a $U(1)_X$ gauge flavor symmetry and enlarging the scalar
sector, a mechanism and an explicit model able to generate
radiatively the hierarchical spectrum of charged fermions
masses and CKM mixing angles is presented. 
The horizontal charge assignment to
particles is such that  no new fermions (beyond the known three
generations of quarks and leptons) are needed. Also, at tree level only 
the t and b quarks and $\tau$ lepton get masses, and to generate radiatively 
the masses for the 
light families, some new exotic scalars are introduced. All of these new 
scalars are electrically charged, so they can not 
acquire VEV as is required in the loop graphs.

The numerical results are presented in Tables \ref{fritzsch:res}, 
\ref{fusaoka:res},
\ref{maslep:fus}, \ref{Tmom} and \ref{tab-res-1}.
Even though the $U(1)_X$ mass scale is not known, the two sets of results 
for the quarks case
do not differ by much and they agree fairly well with the experimental
values, meaning that the mass scale associated with the horizontal symmetry may
be in the range $100 GeV < M_X< 1.0 TeV$, at the same time no experimental bound on
rare decays or AMM is violated. 

A closer look to the analysis shows that the quark mass 
hierarchy in being translated to
the quotient  $v_1/v_2$ which is the hierarchy between the electroweak mass 
scale and the Horizontal $U(1)_X$ mass scale. 
In this way the viability that new physics exists at the electroweak mass scale, or just 
above it, is shown and may help to explain the long-lasting puzzle of the 
enormous range of quark masses and mixing angles.

Since quarks carry baryon number $B=1/3$, the color sextet scalars must 
have $B=-2/3$ (the scalar singlets $\phi_1$ and
$\phi_2$ have B=0); in this way $L_{Y_M}$ is not only $U(1)_X$ invariant 
but conserves color and baryon number as well. On the other hand, $V(\phi)$ 
does not conserve baryon number (the term
$\lambda_4\phi_5\phi_6\phi_7\phi_2$ violates
baryon number).
%and could induce neutron-antineutron oscillations. A
%roughly
%estimate of such oscillations shows that they are proportional to
%$v_2(M^2_{\phi_5}M^2_{\phi_6}M^2_{\phi_7})^{-1}\sim 10^{-19} GeV$ which 
%is negligible in principle. 
Anyway, in the worst of the situations, since 
the offending term  does not enter in the mass matrix for the Higgs scalars,
it may be removed in more realistic models by the introduction of a discrete 
symmetry.

The results are encouraging; even under the assumption that the CKM matrix
is real, and without knowing exactly the $U(1)_X$ mass scale, the
numerical predictions are in the ballpark, implying also a value of order
10 for the Yukawa coupling $Y_Q$, and masses for the exotic scalars being
of order $10^3$ TeV. Thus this model presents a clear mechanism able to
explain the mass hierarchy and mixing of the fermionic fields.

Finally, in the work presented here, the Higgs scalar
used to produce the SSB of the SM gauge group down to $SU(3)_C\times
U(1)_Q$ has zero horizontal charge, and as a consequence the standard $Z$
boson does not mix with the horizontal counterpart. 

\begin{acknowledgments}

One of the authors (A. Z.) acknowledges the hospitality of the theory 
group at CERN and useful conversations with Marcela Carena and Jean 
Pestieau.
\end{acknowledgments}
% ÚLTIMA MODIFICACIÓN, 22 DE MARZO DE 2002

\end{spacing}%%%%%%%%%%%%%%%%%%%%%

\section{BIBLIOGRAPHY}

\begin{thebibliography}{10}

\bibitem{Masas-RMF} E. Garc\'{\i}a, A. Hern\'andez-Galeana, D. Jaramillo, W. A. Ponce and A. Zepeda. Rev. Mex. Fis. \textbf{48-1}, 32 (2002). hep-ph/0006093.

\bibitem{ma}
X.G.He, R.R.Volkas, and D.D.Wu, Phys. Rev. \textbf{D41}, 1630 (1990);
Ernest Ma, Phys. Rev. Lett. \textbf{64}, 2866 (1990).

\bibitem{discrete} Sandip Pakvasa and Hirotaka Sugawara, Phys. Lett.
\textbf{73B}, 61
(1978); Y. Yamanaka, H. Sugawara and  S. Pakvasa, Phys. Rev.
\textbf{D25}, 1895 (1982); K. S. Babu and X.G. He, Phys. Rev.
\textbf{D36}, 3484 (1987); Ernest Ma, Phys. Rev. Lett. \textbf{B62}, 1228
(1989); Ernest Ma and G.G. Wong, Phys. Rev \textbf{D41}, 992 (1990);
G.G. Wong and W.S. Hou, Phys. Rev. \textbf{D50}, 2962 (1994).

\bibitem{continuous} A. Davidson, M. Koca and K. C. Wali, Phys. Rev. Lett.
\textbf{B43}, 92 (1979); A. Davidson, M. Koca and K. C. Wali, Phys. Rev.
\textbf{D20}, 1195 (1979); A. Sirlin, Phys. Rev. \textbf{D22}, 971
(1980); A. Davidson and K. C. Wali, Phys. Rev. \textbf{D21}, 787 (1980);
G. Degrassi and A. Sirlin, Phys. Rev. \textbf{D40}, 3066 (1989); 
P. Binetruy and P. Ramond, Phys. Lett. \textbf{B350}, 49 (1995);
K. Babu and R. N. Mohapatra, Phys. Rev. \textbf{D43}, 2278 (1991);
M. T. Yamawaki and W. W. Wada, Phys. Rev. \textbf{D43}, 2431 (1991);
D. S. Shaw and R. R. Volkas, Phys. Rev. \textbf{D47}, 241 (1993); T.
Yanagida, Phys. Rev. \textbf{D20}, 2986 (1979); E. Papantonopoulos and G.
Zoupanos, Phys. Lett. \textbf{B110}, 465 (1982);  E. Papantonopoulos and G.
Zoupanos, Z. Phys. \textbf{C16}, 361 (1983).


\bibitem{ZEP1} W. A. Ponce, A. Zepeda and J. M. Mira, Z. Phys.
\textbf{C69}, 683 (1996); W. A. Ponce, L. A. Wills and , A. Zepeda, Z. Phys.
\textbf{C73}, 711 (1997).

\bibitem{NusBib} M. C. Gonzalez-Garcia, M. Maltoni, C. Pe\~na-Garay and J. W. F. Valle, Phys. Rev. D \textbf{63}, 033005 (2001).
M. C. Gonzalez-Garcia and M. Maltoni, hep-ph/0202218.

\bibitem{pdg} Particle Data Group, D.E. Groom et al., Eur. Phys. J. 
\textbf{C15},1 (2000).

%\bibitem{pdg} Particle Data Book, C.Caso {\it et all}, Eur. Phys. Journal 
%\textbf{C3}, 1 (1998).

\bibitem{nari}
S. Narison, Nucl. Phys. Proc. Suppl. \textbf{86}, 242-254, (2000), hep-ph/9911454, and references therein.

\bibitem{fusaoka}
H. Fusaoka and Y.Koide, Phys. Rev. \textbf{D57}, 3986 (1998).

\bibitem{fritzsch}
H. Fritzsch and Z. Xing, Prog. Part. Nucl. Phys. \textbf{45}, 1-81 (2000), hep-ph/9912358.

\bibitem{minuit} 
MINUIT, Function Minimization and Error Analysis. CERN
Program Library Long Writeup D506, CERN Geneva, Switzerland.

\bibitem{langacker}J. Erler and P. Langacker, Phys. Rev. Lett. 
\textbf{84}, 212-215 (2000), hep-ph/9910315. P. Langacker
and M. Pl\"umacher, Phys. Rev. D \textbf{62}, 013006 (2000), hep-ph/0001204.

\bibitem{spearman} A.D. Martin, T.D. Spearman, Elementary Particle Theory,
Wiley Interscience Division, 1970.

\bibitem{lapack} LAPACK Users' Guide, E. Anderson, Z. Bai, C. Bischof, J. 
Demmel, J. Dongarra, J. Du Croz, A. Greenbaum, S. Hammarling, A. McKenney,
S. Ostrouchor, D. Sorensen. Society for Industrial and Applied
Mathematics.

\bibitem{BLee} B. W. Lee and R. E. Schrock, Phys. Rev. D {\bf 16}, 1444 
(1977).

\end{thebibliography}
\end{document}